\documentclass[fp,twocolumn]{jpsj3}
\usepackage{txfonts}
\usepackage{graphicx}
\usepackage{dcolumn}
\usepackage{bm}
\usepackage[dvipsnames]{xcolor}

\title{Ground-State Phase Diagram of the Effective Model with Antiferromagnetic Nearest-Neighbor and Ferromagnetic Next-Nearest-Neighbor Interactions in 1/1 Periodic Approximants to Icosahedral Quasicrystals}

\author{Shinji Watanabe$^{*}$}
\inst{Department of Basic Sciences, Kyushu Institute of Technology, Kitakyushu, Fukuoka 804-8550, Japan}

\abst{
By performing 
complete enumeration of all configurations of magnetic moments, 
we have determined the ground-state phase diagram of the effective model with antiferromagnetic nearest-neighbor interactions $J_1$ and ferromagnetic next-nearest-neighbor interactions $J_2$ on the 1/1 periodic approximant to icosahedral quasicrystal. Depending on the ratio of $J_2/J_1$ and the magnetic anisotropy arising from the crystalline electric field, various non-collinear ferrimagnetic orders and antiferromagnetic orders are shown to be stabilized, whose magnetic space groups are identified as $R\bar{3}$, $C2'/m'$, $Pm'm'2$ and $I_{\rm P}m'\bar{3}'$. We have revealed the topological as well as magnetic property of each ground state with degeneracy, which is useful to understand experimental observations of magnetic structures and domains in the rare-earth-based 1/1 approximants. 
}

\begin{document}

\def\gsim{\mathop {\vtop {\ialign {##\crcr 
$\hfil \displaystyle {>}\hfil $\crcr \noalign {\kern1pt \nointerlineskip } 
$\,\sim$ \crcr \noalign {\kern1pt}}}}\limits}
\def\lsim{\mathop {\vtop {\ialign {##\crcr 
$\hfil \displaystyle {<}\hfil $\crcr \noalign {\kern1pt \nointerlineskip } 
$\,\,\sim$ \crcr \noalign {\kern1pt}}}}\limits}

\maketitle

\section{Introduction}

Quasicrystals (QCs) have no three-dimensional periodicity of atoms but possess the structural order. Since the discovery in 1984~\cite{Shechtman}, the understanding of the lattice structures have deepened~\cite{Tsai,Takakura}. However, the electric states and their physical properties have yet to be fully understood. One of the significant questions has been whether the magnetic long-range order is realized in the QCs~\cite{Fisher1999,Sato2000,Goldman2013,Das2023}.

There exist periodic crystals with the atomic configuration being common to that in the QC, which is referred to as the approximant crystal (AC). In the rare-earth-based ACs, magnetic long range orders such as the ferromagnetic (FM) orders and the antiferromagnetic (AFM) orders were observed in Cd$_6$R (R=Tb, Y, Pr, Nd, Sm, Gd, Dy, Ho, Er, Tm, Yb, and Lu)~\cite{Tamura2010,Mori,Das}, Zn-Sc-Tm~\cite{Jazbec} and Au-SM-R (SM=Al, Ga, Si, Ge, R=Gd, Tb, Dy, Ho, and Ce)~\cite{Ishikawa2018,Sato2019,Hiroto,Nawa2023,Labib2023,Labib_ACAF_2025,Hiroto2013,Hiroto2014,Ishikawa2016,Shiino2022,Gebresenbut,Labib2025,Imura,Sawano,Muro,Labib2024,Suzuki2021}. 
Recently, the AFM as well as FM long-range order has been discovered in Au-In-Eu~\cite{Tamura2025} and Au-Ga-R (R=Tb, Gd~\cite{Tamura2021} and Dy~\cite{Takeuchi2023}) respectively, which have attracted much attention. 

\begin{figure}[bt]
\includegraphics[width=8cm]{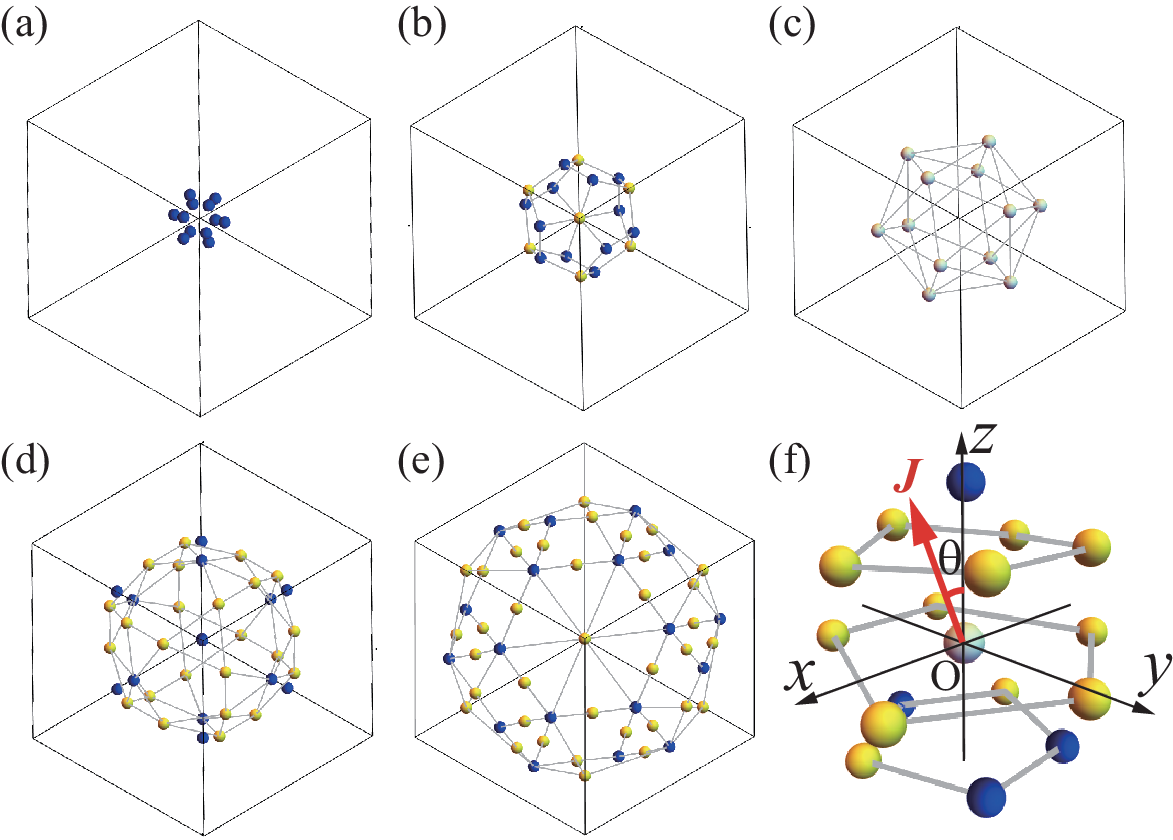}
\caption{(Color online) 
Tsai-type cluster of Au-SM-Tb consists of (a) cluster center, (b) dodecahedron, (c) icosahedron, (d) icosidodecahedron and (e) defect rhombic triacontahedron with Au (yellow), SM (blue), and Tb (gray). 
The frame box denotes the (expanded) unit cell of the bcc lattice of the 1/1 AC Au-SM-Tb. 
(f) Local configuration of atoms surrounding the Tb atom in the Tsai-type cluster. The $z$ axis is the pseudo 5-fold axis. The $yz$ plane is the mirror plane.
The angle $\theta$ is defined between the magnetic easy axis ${\bm J}$ for the CEF ground state and the pseudo 5-fold axis ($z$ axis) within the mirror $(yz)$ plane. 
}
\label{fig:Tsai_cluster}
\end{figure}

These rare-earth-based QCs and ACs are composed of the Tsai-type cluster, which consists of (a) cluster center, (b) dodecahedron, (c) icosahedron, (d) icosidodecahedron and (e) defect rhombic triacontahedron, as shown in Fig.~\ref{fig:Tsai_cluster}. Here we plot the lattice of the 1/1 AC Au$_{70}$Si$_{17}$Tb$_{13}$~\cite{Hiroto} with a lattice constant $a=14.726~{\rm \AA}$ as a typical system (the space group is No. 204 $Im\bar{3}$). 
The rare-earth atoms are occupied at the twelve vertices of the icosahedron [see Fig.~\ref{fig:Tsai_cluster}(c)], where the 4f electrons are responsible for the magnetism. Hence, the rare-earth-based QCs and ACs provide a new research field of strongly correlated electron systems~\cite{NKSato2022}. 
In the 1/1 AC, the icosahedrons are located at the center and corner positions in the body-center-cubic (bcc) lattice. In Fig.~\ref{fig:Tsai_cluster}(f), the local configuration of the rare-earth atom is shown, where the five-fold configuration of ligand atoms exist. As is well known, the 5-fold rotation symmetry is not compatible with the translation symmetry in periodic crystals, well-established crystalline-electric-field (CEF) theory cannot be applied to the present system, which has prevented us from evaluating the CEF microscopically. 

A phenomenological RKKY model assuming the magnetic anisotropy due to the CEF was theoretically studied for the 1/1 AC~\cite{Sugimoto2016}. The RKKY model without the magnetic anisotropy was also studied for the 1/1 AC~\cite{Miyazaki2020,Sugimoto2024}. As for the icosahedral QC, the Monte Carlo simulation for the Heisenberg model has been performed~\cite{WYTF2025}.

Recently, a general formulation of the CEF in the icosahedral QCs and ACs has been developed on the basis of the point-charge model~\cite{WK2021}. 
By applying this framework to the Tb~\cite{WPNAS}-, Dy~\cite{WI2023}-, and Ce~\cite{WFK2026}-based systems, the CEF have been theoretically analyzed, where 
the magnetic easy axis lies in the mirror plane for a wide regime of the CEF parameter in these systems. 
Then, the effective model taking account of the magnetic anisotropy with the nearest neighbor (N.N.) interaction $J_1$ and the next-nearest neighbor interaction $J_2$ was constructed and 
the ground-state phase diagram of the model with the FM interactions ($J_1>0$ and $J_2>0$) ~\cite{WPNAS,WI2026} and the AFM interactions ($J_1<0$ and $J_2<0$)~\cite{WSR,WFK2026} were studied. 

However, the other case that $J_1$ and $J_2$ have different signs e.g., $J_1<0$ and $J_2>0$ is possible in real materials, which have not been studied yet. Hence, in this study, we theoretically analyze the effective model with the AFM interaction for the N.N. bonds and the FM interaction for the N.N.N. bonds. 

The organization of this paper is as follows. In Sect.~\ref{sec:MD}, we introduce our effective model for magnetism in the 1/1 AC and explain the method. In Sect.~\ref{sec:res}, we present our results of the numerical calculations on the basis of the effective model. In Sect.~\ref{sec:disc}, we discuss the characteristic anisotropy angles and the topological Hall effect. In Sect.~\ref{sec:summary}, we summarize our paper.

\section{Model}
\label{sec:MD}

Our study is based on the effective model with the magnetic anisotropy stemming from the CEF in the Au-SM-R~(R=Tb~\cite{WPNAS}, Dy~\cite{WI2023}, and Ce~\cite{WFK2026}). 
\begin{eqnarray}
H=-\sum_{\langle i,j\rangle}J_{ij}\hat{\bm J}_i\cdot\hat{\bm J}_j, 
\label{eq:H}
\end{eqnarray}
where $J_{ij}$ is the interaction between the magnetic moments on the $i$th site and $j$th site 
and $\hat{\bm J}_i$ is the unit vector operator parallel or anti-parallel to the magnetic easy axis on the $i$th site~\cite{WPNAS,WSR}. 
Here, the magnetic easy axis lies in the mirror plane, i.e., the $yz$ plane in the local coordinate shown in Fig.~\ref{fig:Tsai_cluster}(f) and the direction is defined by the anisotropy angle $\theta$ from the pseudo-5-fold axis, i.e., the $z$ axis. 
We employ the atomic coordinate of the 1/1 AC Au$_{70}$Si$_{17}$Tb$_{13}$ with the lattice constant $a=14.726~{\rm \AA}$ (see Fig.~\ref{fig:Tsai_cluster}) as a prototypical system~\cite{Hiroto}. 

\begin{figure}[tb]
\includegraphics[width=7cm]{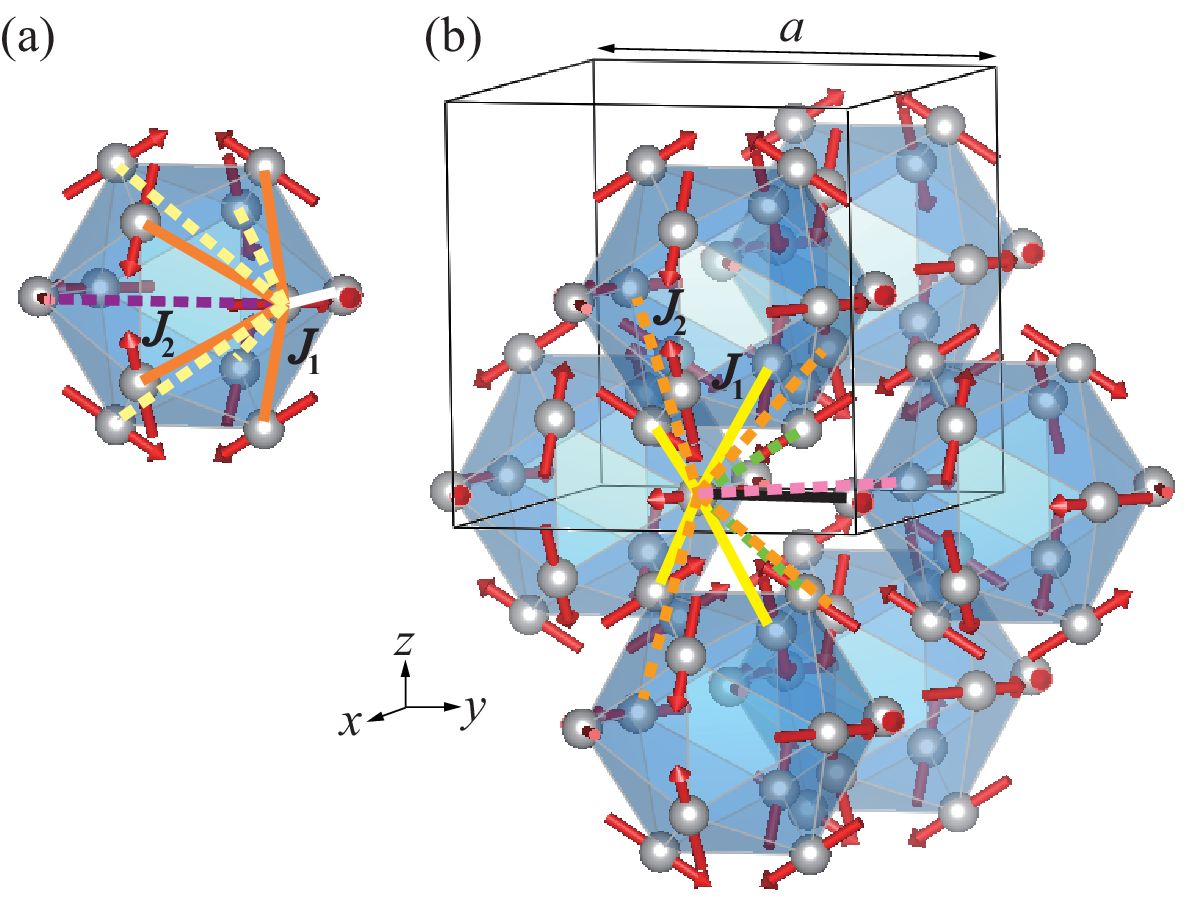}
\caption{(Color online) 
The N.N. interactions $J_1$ and the N.N.N. interactions $J_2$ for (a) intra icosahedron and (b) inter icosahedron are depicted by solid lines and dashed lines, respectively. 
In (b), the unit cell of the bcc lattice of the 1/1 AC with the lattice constant $a$ is denoted as a frame box. 
}
\label{fig:lattice}
\end{figure}

We set the interactions $J_{ij}$ as illustrated in 
Fig.~\ref{fig:lattice}, where the bonds for $J_1$ are depicted as bold solid lines and the bonds for $J_2$ are depicted as bold dashed lines.
Below each bond is explained by noting each color in ( ), as follows. 
As for the intra-icosahedron interaction, 
$J_1$ is set for the 5 bonds for each Tb site with the bond length $0.374a$ (1 bond 
depicted by a white line) and $0.378a$ (4 bonds 
depicted by orange lines) and $J_2$ is set for the 5 bonds for each Tb site with the bond length $0.610a$ (4 bonds 
depicted by yellow dashed lines) and $0.612a$ (1 bond 
depicted by a purple dashed line), as shown in Fig.~\ref{fig:lattice}(a). 
As for the inter-icosahedron interaction, $J_1$ is set for the 5 bonds for each Tb site with the bond length $0.368a$ (4 bonds 
depicted by yellow lines) and $0.388a$ (1 bond 
depicted by a black line) and $J_2$ is set for the 7 bonds with the bond length $0.528a$ (2 bonds 
depicted by green dashed lines), $0.530a$ (4 bonds denoted by orange dashed lines), and $0.539a$ (1 bond 
depicted by a pink dashed line), as shown in Fig.~\ref{fig:lattice}(b).

The bonds can be classified as short bonds with the bond lengths $0.37a\sim 0.38a$ (solid lines in Fig.~\ref{fig:lattice}) and the longer bonds with the bond lengths $0.5a\sim 0.6a$ (dashed lines in Fig.~\ref{fig:lattice}). For simplicity, in this study, we set the interactions for the short bonds as $J_1$ and for the longer bonds as $J_2$ as noted above as a first step of analysis.

In this study, we investigate the ground-state properties of the effective model~(\ref{eq:H}) for $J_1<0$ and $J_2>0$.
By performing 
complete enumeration of all configurations of magnetic moments,  
we have calculated all the energies of the eigenstate of Eq.~(\ref{eq:H}) in the expanded unit cell which contains the two icosahedrons under the periodic boundary condition. Then, we have obtained the ground-state energy. 

To reveal topological property, we calculate the topological charge of the magnetic texture defined on the icosahedron, as follows. 
The topological charge is defined as the solid angle divided by the surface area of the unit sphere~\cite{WPNAS} 
\begin{eqnarray}
n=\frac{\Omega}{4\pi},
\label{eq:tp}
\end{eqnarray}
where $\Omega$ is the solid angle subtended by twelve magnetic moments on the icosahedron 
\begin{eqnarray}
\Omega=\sum_{i,j,k\in {\rm IC}}\Omega_{ijk}.
\label{eq:SI_IC}
\end{eqnarray}
Here, $\Omega_{ijk}$ is the solid angle subtended by three magnetic moments at each triangular surface of the icosahedron (IC), $i$, $j$, and $k$ sites 
\begin{eqnarray}
\Omega_{ijk}=2\tan^{-1}\left[\frac{\chi_{ijk}}{1+{\bm J}_i\cdot{\bm J}_j+{\bm J}_j\cdot{\bm J}_k+{\bm J}_k\cdot{\bm J}_i}\right]
\label{eq:SI_tri}
\end{eqnarray}
\cite{Eriksson}, where ${\bm J}_i$ 
is  
the magnetic-moment vector at the $i$th site in the ground state and 
$\chi_{ijk}$ is the scalar chirality defined by
\begin{eqnarray}
\chi_{ijk}={\bm J}_i\cdot({\bm J}_j\times{\bm J}_k). 
\label{eq:SC}
\end{eqnarray}
%


The total chirality is defined as
\begin{eqnarray}
{\bm \chi}^{\rm T}=\sum_{\langle i,j,k\rangle}\chi_{ijk}\hat{n}_{ijk},
\label{eq:chi_T}
\end{eqnarray}
where the summation is taken over all the three sites 
and $\hat{n}_{ijk}$ denotes the surface normal~\cite{Aoyama2021}. 
The unit  normal vector of the triangle formed by the $i$th, $j$th, and $k$th sites is defined as
\begin{eqnarray}
\hat{n}_{ijk}=\frac{({\bm r}_i-{\bm r}_j)\times({\bm r}_k-{\bm r}_i)}{| ({\bm r}_i-{\bm r}_j)\times({\bm r}_k-{\bm r}_i)|},
\label{eq:nv}
\end{eqnarray}
where ${\bm r}_i$, ${\bm r}_j$, and ${\bm r}_k$ are the position vectors of the $i$th, $j$th, and $k$th sites, respectively. 
The total chirality plays a role as an emergent  
fictitious 
magnetic field, which causes topological Hall effect via the relation~\cite{Tatara}
\begin{eqnarray}
\sigma_{\mu\nu}^{\rm T}\propto\epsilon_{\mu\nu\rho}\chi_{\rho}^{\rm T},
\label{eq:sigma}
\end{eqnarray}
where $\sigma_{\mu\nu}$ is the electric conductivity and $\epsilon_{\mu\nu\rho}$ is the Levi-Civita symbol
\begin{eqnarray}
\epsilon_{\mu\nu\rho}=
\left\{
\begin{array}{rl}
1, & (\mu,\nu,\rho)=(1,2,3), (2,3,1), (3,1,2) \\
-1, & (\mu,\nu,\rho)=(2,1,3), (1,3,2), (3,2,1) \\
0, & {\rm otherwise}. 
\end{array}
\right.
\label{eq:LC}
\end{eqnarray}

According to Ref.~\citen{Tatara}, exponentially decaying factor with respect to the site distance appears in the proportionality coefficient in Eq. (\ref{eq:sigma}). Hence, in the calculation of $\chi_{ijk}$, we consider the triads with short bond lengths with $0.368a$, $0.374a$, or $0.388a$ denoted by solid lines in Fig.~2. Namely, we consider 20 triangular surfaces of the icosahedron with the length of the sides being $0.368a$ and $0.374a$ [see Fig. 2(a)]. We calculate $\chi_{ijk}$ with the order of $i$, $j$, and $k$ being defined in the anticlockwise direction with respect to the normal vector $\hat{n}_{ijk}$ pointing outward from the center of the icosahedron [see Fig.~3(a)]. 

\begin{figure}[tb]
\includegraphics[width=8cm]{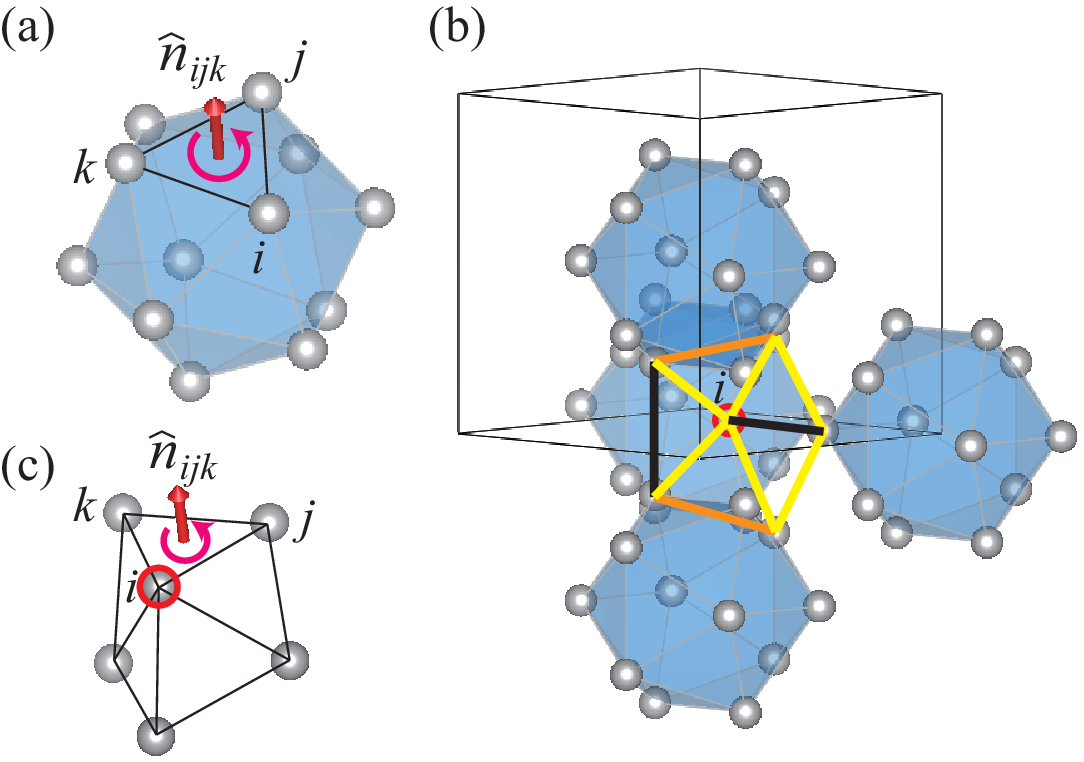}
\caption{(Color online) 
Unit normal vector $\hat{n}_{ijk}$ (red arrow) of the triangular surface formed by the $i$, $j$, and $k$ sites defined by the anticlockwise direction for (a) icosahedron and (b) inter icosahedrons. (c) Pentagonal pyramid composed of the apex and five basal vertices, which correspond to the vertex of the icosahedron and vertices of the outer icosahedrons, respectively. 
}
\label{fig:nv}
\end{figure}

As denoted by yellow and black solid lines in Fig.~\ref{fig:lattice}(b), there exist inter-icosahedron bonds with the short bond length, which form triangles.  As shown in Fig.~\ref{fig:nv}(b), the vertex of the icosahedron denoted by the $i$th site constitutes five triangles, each of which has short side lengths. Namely, each of the five triangles has two sides with the side length $0.368a$ $(0.388a)$ corresponding to the yellow (black) solid lines in Fig.~\ref{fig:lattice}(b) and the extra one side with the side length $0.378a$ or $0.368a$ corresponding to the orange or white solid line respectively in Fig.~\ref{fig:lattice}(a). For the five triangular surfaces, we calculate $\chi_{ijk}$ with the order of $i$, $j$, and $k$ being defined in the anticlockwise direction with respect to the normal vector $\hat{n}_{ijk}$ defined by Eq.~(\ref{eq:nv}) pointing outward from the pentagonal pyramid [see Fig.~\ref{fig:nv}(c)]. Here, the apex of the pyramid depicted by red circle is the vertex of the icosahedron and five basal vertices are the vertices of outer icosahedrons. Since any diagonal line intersecting the basal plane of the pentagonal pyramid is much longer than the bond length $0.368a\sim 0.388a$ denoted by solid lines, we neglected the contribution from the triangles in the basal plane in the calculation of $\chi_{ijk}$. 

In this study, we report the total chirality ${\bm \chi}^{\rm T}$ per icosahedron calculated for intra and inter icosahedrons based on Eq.~(\ref{eq:chi_T}).

The total magnetic moment per icosahedron is defined by
\begin{eqnarray}
{\bm J}_{\rm IC}=\sum_{i=1}^{12}{\bm J}_{i}. 
\label{eq:J_IC}
\end{eqnarray}

We have performed complete enumeration of all configurations of magnetic moments to obtain the ground state of the effective model (\ref{eq:H}) for various $J_1 (<0)$ and $J_2 (>0)$ and the anisotropy angle $0\le\theta\le 180^{\circ}$. 
For each obtained ground state, we have calculated the topological charge $n$ by Eq.~(\ref{eq:tp}), total chirality ${\bm \chi}^{\rm T}$ by Eq.~(\ref{eq:chi_T}), and the total magnetic moment per icosahedron ${\bm J}_{\rm IC}$ by Eq.~(\ref{eq:J_IC}). Then, we have determined the ground-state phase diagram. In the next section, we present the results and discuss the property of each magnetically-ordered phase. 

\section{Results}
\label{sec:res}

\subsection{Ground-state phase diagram}

\begin{figure}[tb]
\includegraphics[width=8cm]{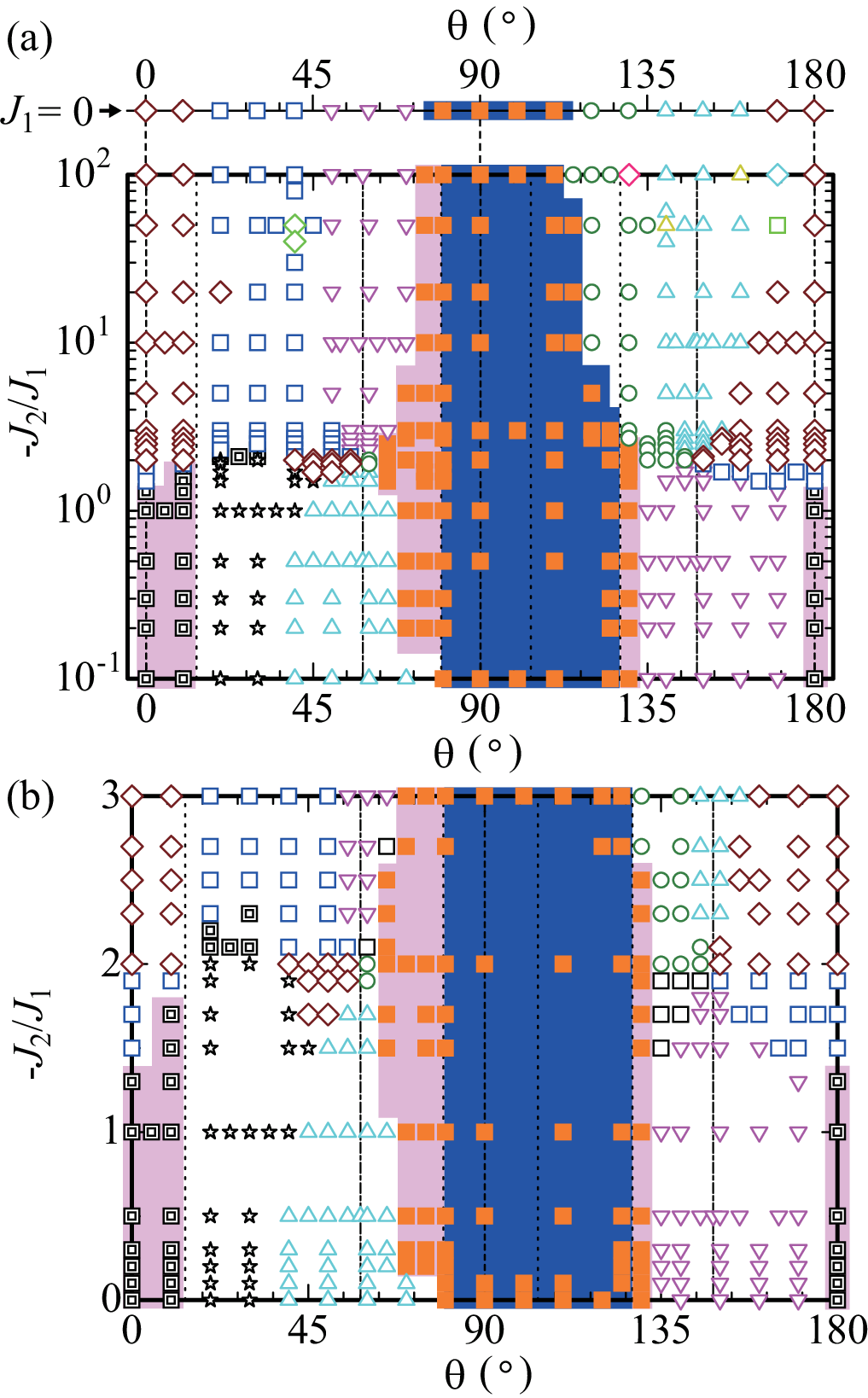}
\caption{(Color online) 
The ground-state phase diagram in the plane of $-J_2/J_1$ for $J_1<0$ and $J_2>0$ and the anisotropy angle $\theta$. 
(a) $J_1=0$ (upper panel) and $10^{-1}\le -J_2/J_1\le 10^2$ in logarithmic plot (lower panel). 
(b) The enlargement for $0\le -J_2/J_1\le 3$. 
In (a) and (b), the colored regions denote magnetic orders characterized by the topological charge $|n|=3$ (blue), $1$ (pink), and $0$ (white). 
The open symbols denote FM orders and the filled symbols denote AFM orders. 
The vertical dashed lines at $\theta=0^{\circ}$, $90^{\circ}$ and $180^{\circ}$ are the guides for the eyes. The vertical dotted lines at $\theta=\theta_{{\rm c}0}=13.570^{\circ}, \theta_{\rm c1}=79.475^{\circ}$, $\theta_{\rm c2}=103.570^{\circ}$ and $\theta_{\rm c3}=127.665^{\circ }$ and the vertical dash-dotted lines at $\theta=\theta_0=58.282^{\circ}$ and $\theta_1=148.282^{\circ}$ are the guides for the eyes (see text). 
}
\label{fig:PD}
\end{figure}

We have determined the ground-state phase diagram of the effective model (\ref{eq:H}) as shown in Fig.~\ref{fig:PD}. 
Depending on the ratio of the N.N. and N.N.N. interactions and the anisotropy angle $\theta$, various non-collinear magnetic structures are stabilized. 
The open symbols denote the FM orders with the magnetic propagation vector $(hkl)=(000)$. We refer to these orders as non-collinear ferrimagnetic orders since  magnetic moments on each icosahedron are aligned non-collinearly, which result in finite magnetization per icosahedron.  
The filled symbol denotes the AFM order with the magnetic propagation vector $(hkl)=(111)$ in ${\bm q}=\frac{2\pi}{a}(hkl)$. 

In Fig.~\ref{fig:PD}(a), the phase diagram for $J_1=0$ is shown in the upper panel. In the lower panel, the phase diagram for $10^{-1}\le-J_2/J_1\le 10^2$ is shown in logarithmic scale. In Fig.~\ref{fig:PD}(b), the phase diagram for $0\le-J_2/J_1\le 3$ is shown. The AFM order is stabilized around $90^{\circ}$, while the ferrimagnetic orders are stabilized in the other regions near $\theta=0^{\circ}$ and $180^{\circ}$. Since the ground state with all the magnetic moments being inverted is energetically degenerate, the ground state at $\theta=180^{\circ}$ is the time-reversal state of the ground state at $\theta=0^{\circ}$. 

We have calculated the topological charge for each magnetic state on the icosahedron on the basis of Eq.~(\ref{eq:tp}) and the magnetic orders characterized by $|n|=3$,  
$1$, and $0$ are depicted by the regions colored in blue, pink, and white, respectively. 
As discussed in Sect.~\ref{sec:disc}, there are characteristic anisotropy angles $\theta_{{\rm c}i}$ $(i=0, 1, 2, 3)$ defined by 
$\theta_{{\rm c}0}=13.570\dots^{\circ}$, $\theta_{{\rm c}1}=79.475\dots^{\circ}$, $\theta_{{\rm c}2}=103.570\dots^{\circ}$, and $\theta_{{\rm c}3}=127.665\dots^{\circ}$. We plot $\theta_{{\rm c}i}$ as the vertical dashed lines in Fig.~\ref{fig:PD}. 

As for the ferrimagnetic orders denoted by the pink-inverted triangle and the light-blue triangle, there also exist the characteristic anisotropy angles $\theta_0=58.282\dots^{\circ}$ and $\theta_1=148.282\dots^{\circ}$ as explained below, which are plotted as the dash-dotted lines in Fig.~\ref{fig:PD}. 

In the following subsections, each property of the ferrimagnetic orders and the AFM order is explained.

\subsection{Ferrimagnetic orders}

\begin{figure}[tb]
\includegraphics[width=6.5cm]{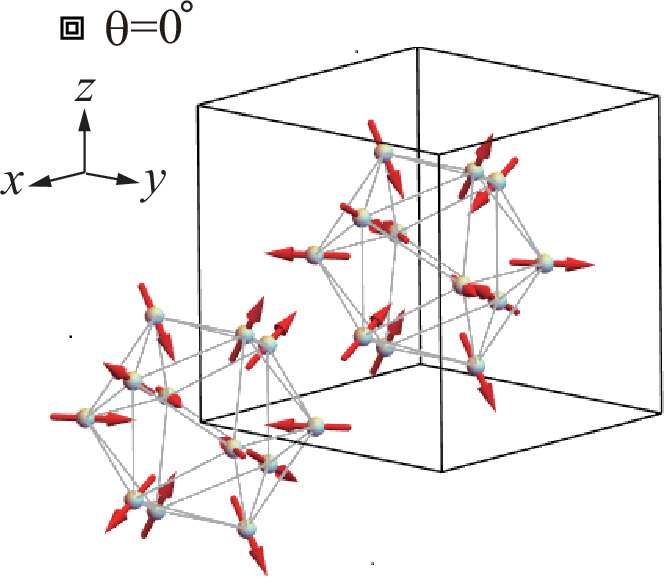}
\caption{(Color online) 
The non-collinear ferrimagnetic state for $\theta=0^{\circ}$ denoted by the black double-square symbol in Fig.~\ref{fig:PD}. The framebox denotes the unit cell. 
}
\label{fig:blck_db_sq}
\end{figure}

The black double-square symbol in Fig.~\ref{fig:PD} denotes the ferrimagnetic order as shown in Fig.~\ref{fig:blck_db_sq}, where the magnetic state on the corner icosahedron is distributed at all corners of the unit cell depicted as a frame box, one of which is illustrated as a representative one. 
The magnetic space group is $Pm'm'2$ (No.25.4.158 in the OG notation~\cite{OG} and No.25.60 in the BNS notation~\cite{BNS}). 
Namely, the magnetic moments in Fig.~\ref{fig:blck_db_sq} are invariant under the mirror operation with respect to the mirror plane perpendicular to the $x$ as well as the $z$ axis with time-reversal operation. The magnetic moments are also invariant under the 2-fold-axis rotation around the $y$ axis. 

The total magnetic moments per the center icosahedron and the corner icosahedrons point to the 2-fold-axis direction. For the magnetic state shown in Fig.~\ref{fig:blck_db_sq}, each icosahedron in the unit cell exhibits ${\bm J}_{\rm IC}=(0.00, 2.10, 0.00)$, which points to the positive-$y$ direction. 
The total chirality points to the positive-$y$ direction as ${\bm \chi}^{\rm T}=(0.00,1.12,0.00)$ for each icosahedron in the unit cell. 
Our calculation based on Eq.~(\ref{eq:tp}) shows that the topological charge $n$ is $-1$ for the magnetic state on the center and corner icosahedrons in Fig.~\ref{fig:blck_db_sq}.
The degeneracy of this ground state is 12 for $-J_2/J_1>0$. 
From cubic $Im\bar{3}$ to primitive $Pm'm'2$, the way of choosing the unique axis, e.g., $y$ axis among three 2-fold axes is three. Then, the way of choosing $x$, $z$ axis between extra two axes is two, which results in $3\times 2=6$. This six degree and time reversal one yield $6\times 2=12$ degree. Our numerical calculation confirmed this as the number of degeneracy, 12 of the ground states, which corresponds to the number of the magnetic domains. 
It is noted here that for $J_2=0$, the degeneracy is 192, where the 12 degenerate states with the magnetic state presented in Fig.~\ref{fig:blck_db_sq} are included. 

These characteristics are summarized in Table~\ref{tb:mag_state}, where 
the independent magnetic position normalized by the lattice constant is listed in the 4th column. We also list ${\bm J}_{\rm IC}$, ${\bm \chi}^{\rm T}$, and the absolute value of the topological charge $|n|$ in the 6th, 7th, and 8th column for each degenerate state, respectively. Note that since time-reversal state is energetically degenerate, there exists the ground state with all the components of ${\bm J}_{\rm IC}$ and ${\bm \chi}^{\rm T}$ having opposite signs, which is not listed in Table~\ref{tb:mag_state}. The same is for the other ordered states below. 

\begin{figure}[tb]
\includegraphics[width=8cm]{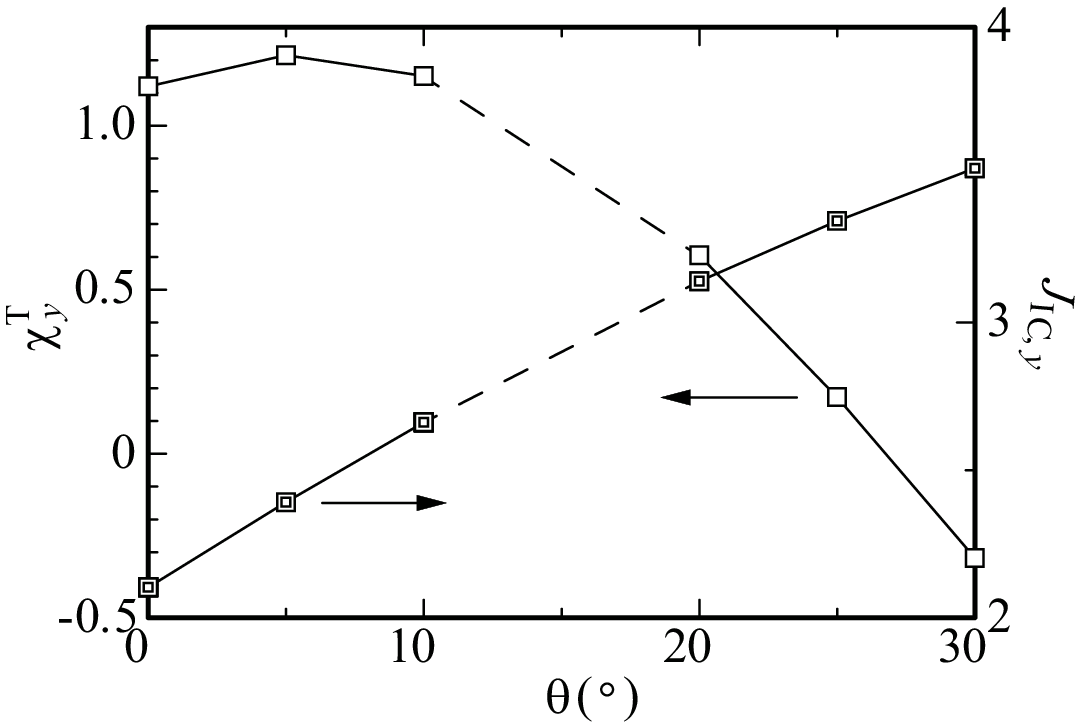}
\caption{
The anisotropy angle $\theta$ dependences of the total magnetic moment per icosahedron $J_{{\rm IC},y}$ (double square, right axis) and the total chirality $\chi^{\rm T}_{y}$ (square, left axis) for the ferrimagnetic state denoted by the double-square symbol in Fig.~\ref{fig:PD}. 
The dashed lines between the two ordered regimes are the guides for the eyes. 
} 
\label{fig:blck_db_sq_J_chiT}
\end{figure}

This ferrimagnetic order shown in Fig.~\ref{fig:blck_db_sq} shares the regions for $0^{\circ}\le\theta\lsim 10^{\circ}$ as well as $20^{\circ}\lsim\theta\lsim 30^{\circ}$ in the phase diagram shown in Fig.~\ref{fig:PD}. The anisotropy angle $\theta$ dependences of the total magnetic moment per icosahedron $J_{{\rm IC},y}$ and the total chirality $\chi^{\rm T}_y$ are shown in Fig.~\ref{fig:blck_db_sq_J_chiT}.

\begin{figure}[tb]
\includegraphics[width=6.5cm]{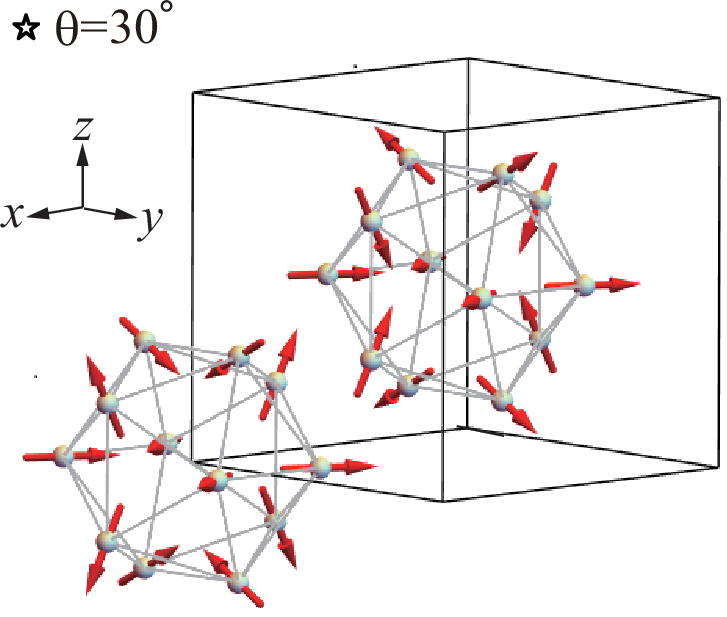}
\caption{(Color online) 
The non-collinear ferrimagnetic state for $\theta=30^{\circ}$ denoted by the black-star symbol in Fig.~\ref{fig:PD}. The frame box denotes the unit cell.
}
\label{fig:blck_str}
\end{figure}

The black-star symbol in Fig.~\ref{fig:PD} denotes the ferrimagnetic order as shown in Fig.~\ref{fig:blck_str}, where the magnetic state on the corner icosahedron is distributed at all corners of the unit cell depicted as a frame box, one of which is illustrated as a representative one. 
The magnetic space group is $Pm'm'2$ (No.25.4.158 in the OG notation~\cite{OG} and No.25.60 in the BNS notation~\cite{BNS}). Namely, the magnetic moments in Fig.~\ref{fig:blck_str} are invariant under the mirror operation with respect to the mirror plane perpendicular to the $x$ as well as $z$ axis with time-reversal operation. The magnetic moments are also invariant under the 2-fold axis rotation around the $y$ axis. 

The total magnetic moments per the center icosahedron and the corner icosahedrons are directed along the 2-fold axis. For the magnetic state shown in Fig.~\ref{fig:blck_str}, each icosahedron in the unit cell exhibits ${\bm J}_{\rm IC}=(0.00, 1.90, 0.00)$, which points to the positive-$y$ direction. 
The total chirality points to the positive-$y$ direction as ${\bm \chi}^{\rm T}=(0.00,0.28,0.00)$ for each icosahedron in the unit cell. 
The topological charge $n$ is zero for the magnetic state on the center and corner icosahedrons in Fig.~\ref{fig:blck_str}.
The degeneracy of this ground state is 12 for $-J_2/J_1>0$. 
The degeneracy corresponds to the number of magnetic domains which arises from lowering symmetry from $Im\bar{3}$ to $Pm'm'2$, as noted above. 
It is noted here that for $J_2=0$, the degeneracy is 192, where the 12 degenerate states with the magnetic state presented in Fig.~\ref{fig:blck_str} are included. 
These characteristics are summarized in Table~\ref{tb:mag_state}. 

\begin{figure}[tb]
\includegraphics[width=8cm]{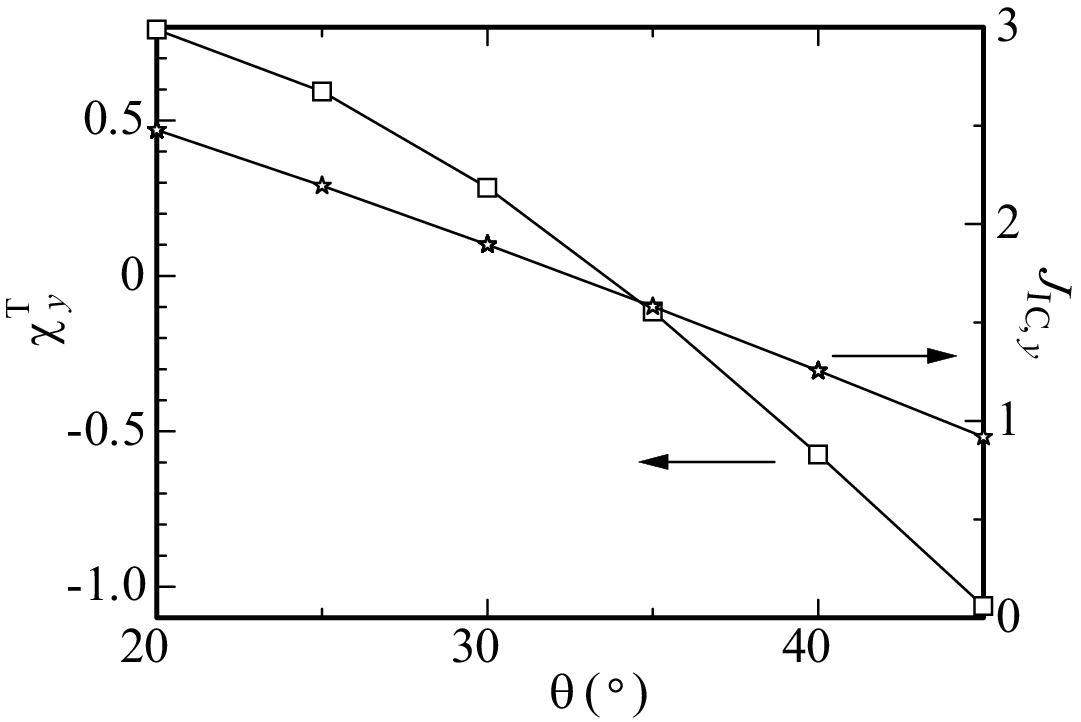}
\caption{
The anisotropy angle $\theta$ dependences of the total magnetic moment per icosahedron $J_{{\rm IC},y}$ (star, right axis) and the total chirality $\chi^{\rm T}_{y}$ (square, left axis) for the ferrimagnetic state 
denoted by the black-star symbol in Fig.~\ref{fig:PD}. 
} 
\label{fig:blck_str_J_chiT}
\end{figure}

This ferrimagnetic order shown in Fig.~\ref{fig:blck_str} shares the region for $20^{\circ}\lsim\theta\lsim 45^{\circ}$ in the phase diagram presented in Fig.~\ref{fig:PD}. The anisotropy angle $\theta$ dependences of the total magnetic moment per icosahedron $J_{{\rm IC},y}$ and the total chirality $\chi^{\rm T}_y$ are shown in Fig.~\ref{fig:blck_str_J_chiT}.

\begin{figure}[tb]
\includegraphics[width=8cm]{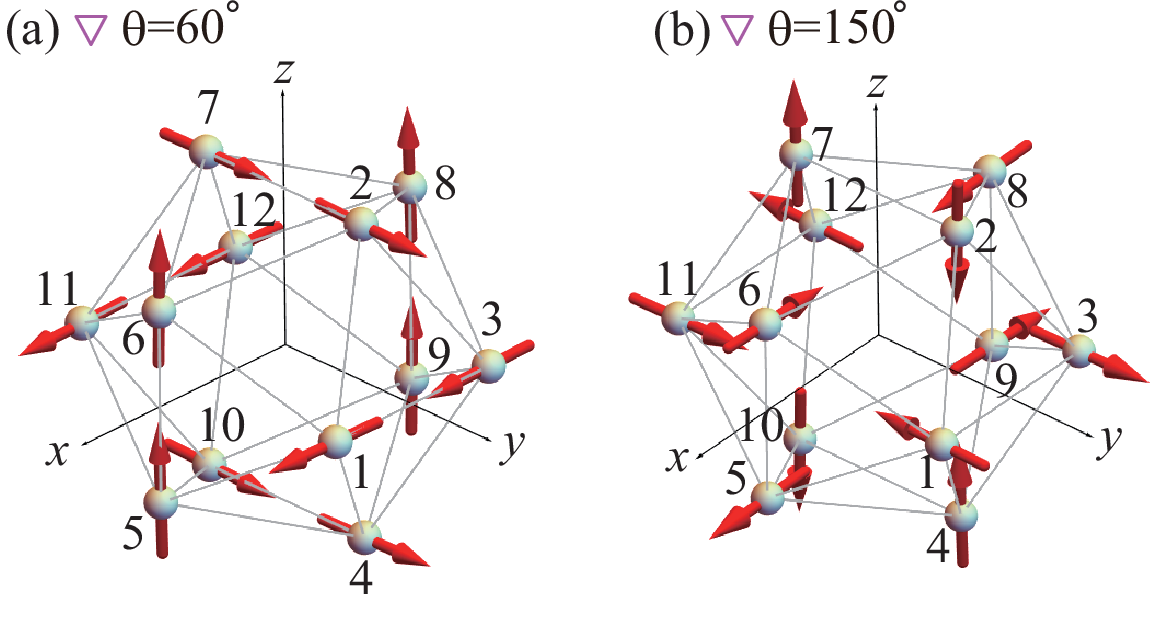}
\caption{(Color online) 
The non-collinear ferrimagnetic state for (a) $\theta=60^{\circ}$ and (b) $\theta=150^{\circ}$ denoted by the pink-inverted-triangle symbol in Fig.~\ref{fig:PD}. In (a) and (b), the number labels the rare-earth site.
}
\label{fig:pink_inv_tri}
\end{figure}

The pink-inverted-triangle symbol in Fig.~\ref{fig:PD} denotes the ferrimagnetic order as shown in Fig.~\ref{fig:pink_inv_tri}, where the cases for the anisotropy angle $\theta=60^{\circ}$ and $\theta=150^{\circ}$ are illustrated in Fig.~\ref{fig:pink_inv_tri}(a) and Fig.~\ref{fig:pink_inv_tri}(b), respectively. 
The magnetic space group is $R\bar{3}$ (No. 148.1.1247 in OG notation~\cite{OG} and 148.17 $R\bar{3}$ in BNS notation~\cite{BNS}). Figures~\ref{fig:pink_inv_tri}(a) and \ref{fig:pink_inv_tri}(b) are the views from the 3-fold axis, i.e., the (111) direction. The total magnetic moment per icosahedron points to the principal-axis direction of rhombohedral $R\bar{3}$, e.g., the (111) direction in the magnetic states for Figs~\ref{fig:pink_inv_tri}(a) and \ref{fig:pink_inv_tri}(b). The total chirality points to the opposite direction of the principal axis of rhombohedral $R\bar{3}$, e.g., the $-(111)$ direction. The degeneracy of this ground state is 8 for $-J_2/J_1>0$. 
The way of choosing the one 3-fold axis for rhombohedral lattice from cubic $Im\bar{3}$ is 4, i.e., four diagonal directions of cubic. This four degree and time reversal one give $4\times 2=8$. Our numerical calculation confirmed this as the number of degeneracy, 8 of the ground states, which corresponds to the number of the magnetic domains.~\cite{WI2026}  
For $J_2=0$, the degeneracy is 64, where the 8 degenerated states with that presented in Fig.~\ref{fig:pink_inv_tri}(b) are included. These characteristics are summarized in Table~\ref{tb:mag_state}.

\begin{figure}[tb]
\includegraphics[width=8cm]{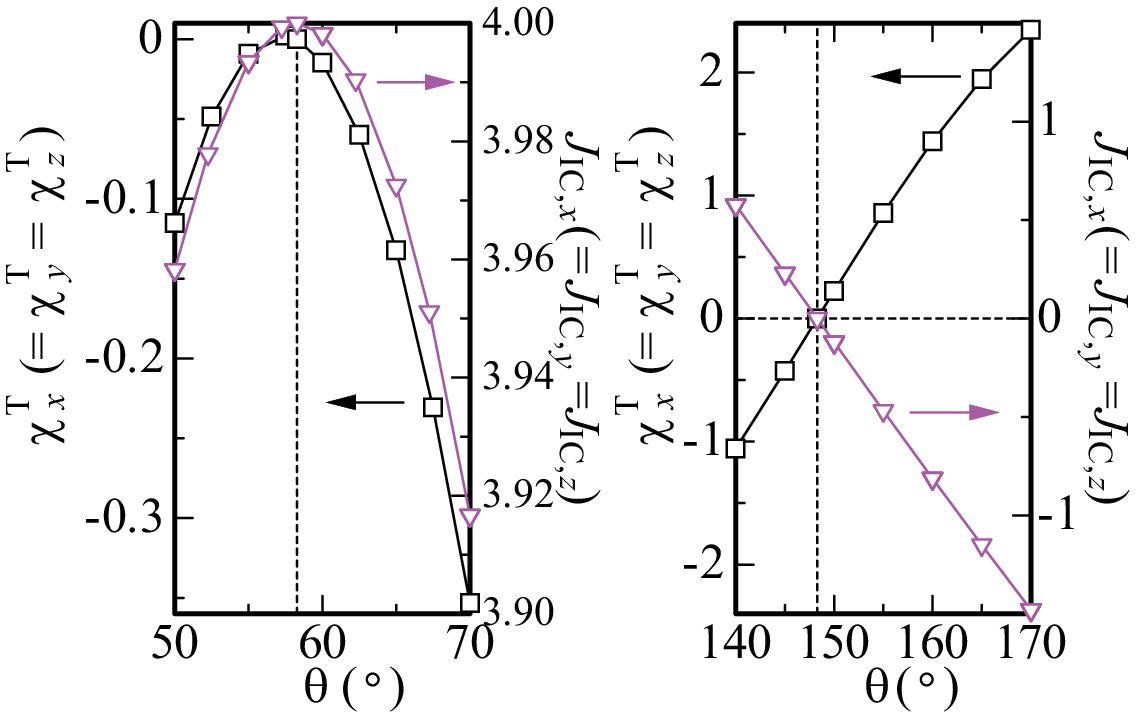}
\caption{(Color online) 
The anisotropy angle $\theta$ dependences of the total magnetic moment per icosahedron (inverted triangle, right axis) and the total chirality (square, left axis) for the ferrimagnetic state 
denoted by the pink-inverted-triangle symbol in Fig.~\ref{fig:PD}. In (a) and (b), the vertical dashed line at $\theta=\theta_0$ and $\theta=\theta_1$ are the guide for the eyes, respectively (see text). 
}
\label{fig:pink_inv_tri_J_chiT}
\end{figure}

This ferrimagnetic order shares the regions for $50^{\circ}\lsim\theta\lsim 70^{\circ}$ and $-J_2/J_1\gsim 2.3$ as well as for $140^{\circ}\lsim\theta\lsim 170^{\circ}$ and $0\le -J_2/J_1\lsim 1.8$ in the phase diagram presented in Fig.~\ref{fig:PD}. The anisotropy angle dependences of the total magnetic moment per icosahedron $J_{{\rm IC}, x} (=J_{{\rm IC},y}=J_{{\rm IC},z})$ and the total chirality $\chi^{\rm T}_x (=\chi^{\rm T}_y=\chi^{\rm T}_z)$ are shown in Figs.~\ref{fig:pink_inv_tri_J_chiT}(a) and \ref{fig:pink_inv_tri_J_chiT}(b), respectively. 
In Fig.~\ref{fig:pink_inv_tri_J_chiT}(a), $J_{{\rm IC},\mu}$ and $\chi^{\rm T}$ $(\mu=x, y, z)$ have the maximum at $\theta=\theta_0\equiv{\tan}^{-1}(\tau)=58.28254\dots^{\circ}$~\cite{W2026}, where $\tau$ is the golden mean $\tau=(1+\sqrt{5})/2$. 

At $\theta=\theta_0$, the magnetic moments at the $i=1$st, 3rd, 11th, and 12th sites in Fig.~\ref{fig:pink_inv_tri}(a) point to the $x$-axis direction. Moreover, the magnetic moments at the $i=2$nd, 4th, 7th, and 
10th  
sites point to the $y$-axis direction and those at the $i=5$, 6, 8, and 9th sites point to the $z$-axis direction~\cite{WI2026}. Hence, $J_{{\rm IC},\mu}$ $(\mu=x, y, z)$ takes the maximum at $\theta=\theta_0$ in Fig.~\ref{fig:pink_inv_tri_J_chiT}(a). 

In Fig.~\ref{fig:pink_inv_tri_J_chiT}(b), 
as $\theta$ increases, $J_{{\rm IC}, \mu}$ $(\mu=x, y, z)$ decreases monotonically with the sign being changed at $\theta=\theta_1\equiv{\tan}^{-1}(\tau)+90^{\circ}=148.28254\dots^{\circ}$~\cite{W2026}. As $\theta$ increases, each component of the total chirality $\chi^{\rm T}_{\mu}$ $(\mu=x,y,z)$ increases with the sign being changed at $\theta=\theta_1$. 

At $\theta=\theta_1$, the magnetic moments at the $i=1$st and 12th sites as well as the $i=3$rd and 11th sites in Fig.~\ref{fig:pink_inv_tri}(b) are aligned collinearly, the former of which are anti-parallel to the latter. Similarly, the magnetic moments at the $i=2$nd and $10$th sites as well as the 4th and 7th sites are aligned collinearly, the former of which are anti-parallel to the latter and those at the $i=5$th and 8th sites as well as the 6th and 9th sites are aligned colinearly, the former of which are anti-parallel to the latter. Hence, $J_{{\rm IC},\mu}$ $(\mu=x, y, z)$ becomes zero at $\theta=\theta_1$ in Fig.~\ref{fig:pink_inv_tri_J_chiT}(b).

At $\theta=\theta_1$, it is easy to understand why the state denoted by the pink-inverted triangle symbol is realized as the ground state for $J_2=0$. Focusing on the $i=1$st site [see Fig.~\ref{fig:pink_inv_tri}(b)], the inner products $\hat{\bm J}_i\cdot\hat{\bm J}_j$ with the short-bond pairs i.e., the $j=2$, 3, 4, 5 and $6$th sites for intra icosahedron are zero except for the $j=3$rd site, which gives the minimized value $\hat{\bm J}_1\cdot\hat{\bm J}_3=-1$. Since any magnetic states at the $j=2$, 4, 5 and 6th sites are orthogonal to $\hat{\bm J}_1$, this state has the lowest energy for the AFM interaction $J_1<0$. As for the short-bond pairs for inter icosahedrons, i.e., $j=7$, 8, 9, 10 and 11th sites, the inner products $\hat{\bm J}_1\cdot\hat{\bm J}_j$ are zero except for the $j=11$th site, which gives the minimized value $\hat{\bm J}_1\cdot\hat{\bm J}_{11}=-1$. Since any magnetic states at the $j=7$, 8, 9 and 10th sites are orthogonal to $\hat{\bm J}_1$, this state has the lowest energy for $J_1<0$. This argument can be applied to all the other $i=2$, $3, \cdots, 12$th sites in the similar way. Hence, it turns out that the state denoted by the pink-inverted triangle symbol is realized as the ground at $\theta=\theta_1$ for the AFM interaction $J_1$ and $J_2=0$ in Fig.~\ref{fig:PD}(b).

It should be noted that the ferrimagnetic order denoted by the pink-inverted-triangle symbol was experimentally observed in the 1/1 AC Au-Si-Tb~\cite{Hiroto,Labib2023,Gebresenbut}.

\begin{figure}[tb]
\includegraphics[width=8cm]{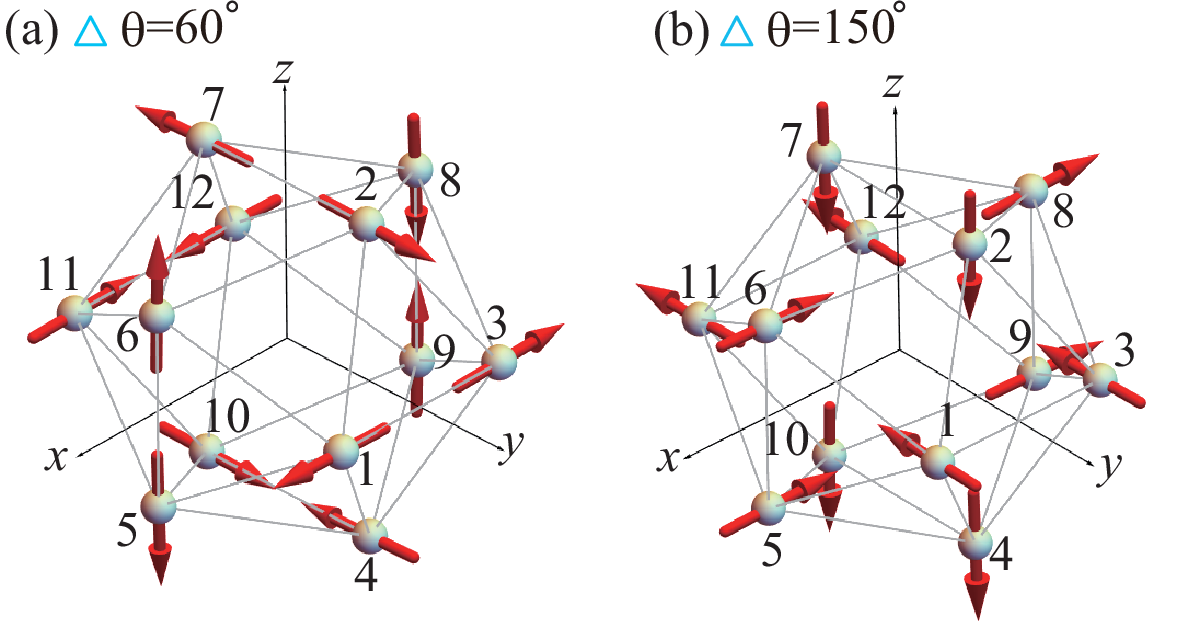}
\caption{(Color online) 
The non-collinear ferrimagnetic state for (a) $\theta=60^{\circ}$ and (b) $\theta=150^{\circ}$ denoted by the light-blue triangle symbol in Fig.~\ref{fig:PD}. In (a) and (b), the number labels the rare-earth site.
}
\label{fig:lbl_tri}
\end{figure}

The light-blue-triangle symbol in Fig.~\ref{fig:PD} denotes the ferrimagnetic order as shown in Fig.~\ref{fig:lbl_tri}, where the cases for the anisotropy angle $\theta=60^{\circ}$ and $\theta=150^{\circ}$ are illustrated in Fig.~\ref{fig:lbl_tri}(a) and Fig.~\ref{fig:lbl_tri}(b), respectively. The magnetic space group is $R\bar{3}$ (No. 148.1.1247 in OG notation~\cite{OG} and 148.17 $R\bar{3}$ in BNS notation~\cite{BNS}). Figures~\ref{fig:lbl_tri}(a) and \ref{fig:lbl_tri}(b) are the views from the 3-fold axis, i.e., the (111) direction. The total magnetic moment per icosahedron is directed oppositely to the principal axis of rhombohedral $R\bar{3}$, e.g., the $-(111)$ direction in the magnetic states for Figs~\ref{fig:lbl_tri}(a) and \ref{fig:lbl_tri}(b). The total chirality points to the principal-axis direction of rhombohedral $R\bar{3}$, e.g., the -(111) direction. The degeneracy of this ground state is 8 for $-J_2/J_1>0$. 
The degeneracy corresponds to the number of magnetic domains which arises from lowering symmetry from $Im\bar{3}$ to $R\bar{3}$, as noted above. 
For $J_2=0$, the degeneracy is 64, where the 8 degenerated states with that presented in Fig.~\ref{fig:lbl_tri}(a) are included. These characteristics are summarized in Table~\ref{tb:mag_state}. 

\begin{figure}[tb]
\includegraphics[width=8cm]{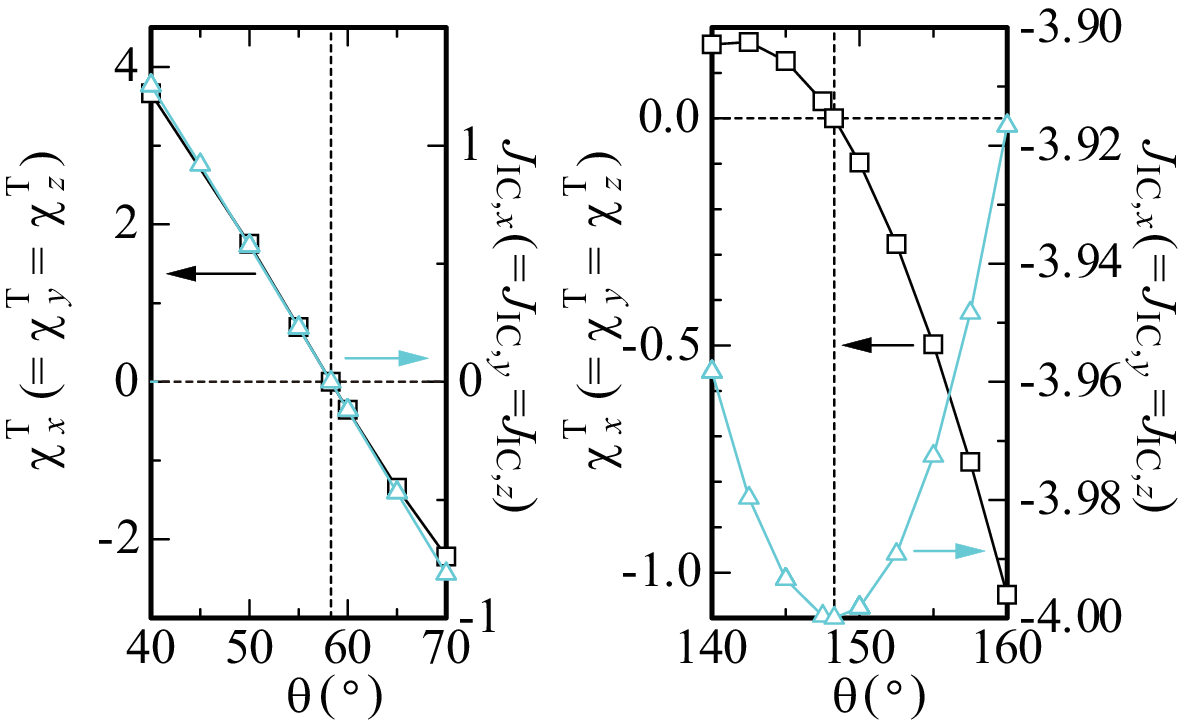}
\caption{(Color online) 
The anisotropy angle $\theta$ dependence of the total magnetic moment per icosahedron (light-blue triangle, right axis) and the total chirality (square, left axis) for the ferrimagnetic state 
denoted by the light-blue-triangle symbol in Fig.~\ref{fig:PD}. In (a) and (b), the vertical dashed lines at $\theta=\theta_0$ and $\theta=\theta_1$ are the guide for the eyes, respectively (see text). 
}
\label{fig:lbl_tri_J_chiT}
\end{figure}

This ferrimagnetic order shown in Figs.~\ref{fig:lbl_tri}(a) and \ref{fig:lbl_tri}(b) shares the regions for $40^{\circ}\lsim\theta\lsim 70^{\circ}$ and $0\le -J_2/J_1\lsim 1.7$ as well as for $140^{\circ}\lsim\theta\lsim 160^{\circ}$ and $0\le -J_2/J_1\lsim 2.3$ in the phase diagram presented in Fig.~\ref{fig:PD}. The anisotropy angle dependences of the total magnetic moment per icosahedron $J_{{\rm IC}, x} (=J_{{\rm IC},y}=J_{{\rm IC},z})$ and the total chirality $\chi^{\rm T}_x (=\chi^{\rm T}_y=\chi^{\rm T}_z)$ are shown in Figs.~\ref{fig:lbl_tri_J_chiT}(a) and \ref{fig:lbl_tri_J_chiT}(b), respectively. In Fig.~\ref{fig:lbl_tri_J_chiT}(a), as $\theta$ increases, $J_{{\rm IC}, \mu}$ $(\mu=x, y, z)$ decreases monotonically with the sign being changed at $\theta=\theta_0$. Each component of the total chirality $\chi^{\rm T}_{\mu}$ $(\mu=x,y,z)$ decreases as $\theta$ increases and becomes zero at $\theta=\theta_0$. As $\theta$ further increases, $\chi^{\rm T}_{\mu}$ decreases. 

At $\theta=\theta_0$, the magnetic moments at the $i=1$st and 12th sites as well as the $i=3$rd and 11th sites in Fig.~\ref{fig:lbl_tri}(a) are aligned collinearly, the former of which are anti-parallel to the latter. Similarly, the magnetic moments at the $i=2$nd and $10$th sites as well as the 4th and 7th sites are aligned collinearly, the former of which are anti-parallel to the latter and those at the $i=5$th and 8th sites as well as the 6th and 9th sites are aligned colinearly, the former of which are anti-parallel to the latter. Hence, $J_{{\rm IC},\mu}$ $(\mu=x, y, z)$ becomes zero at $\theta=\theta_0$ in Fig.~\ref{fig:lbl_tri_J_chiT}(a).

In Fig.~\ref{fig:lbl_tri_J_chiT}(b), as $\theta$ increases, $\chi^{\rm T}_{\mu}$ decreases monotonically with the sign being changed at $\theta=\theta_1$, while $J_{{\rm IC},\mu}$ $(\mu=x, y, z)$ shows a concave $\theta$-dependence with the minimum at $\theta=\theta_1$. 
At $\theta=\theta_1$, the magnetic moments at the $i=1$st, 3rd, 11th, and 12th sites in Fig.~\ref{fig:lbl_tri}(b) point to the $-x$ direction. Moreover, the magnetic moments at the $i=2$nd, 4th, 7th, 
10th  
 sites point to the $-y$ direction and those at the $i=5$, 6, 8, and 9th sites point to the $-z$ direction~\cite{WI2026}. Hence, $J_{{\rm IC},\mu}$ $(\mu=x, y, z)$ takes the minimum at $\theta=\theta_1$ in Fig.~\ref{fig:lbl_tri_J_chiT}(b). 
 
At $\theta=\theta_0$, it is understandable why the state denoted by the light-blue-triangle symbol is realized as the ground state for $J_2=0$. Focusing on the $i=1$st site [see Fig.~\ref{fig:lbl_tri}(a)], the inner products $\hat{\bm J}_i\cdot\hat{\bm J}_j$ with the short-bond pairs i.e., the $j=2$, 3, 4, 5 and $6$th sites for intra icosahedron  [see solid  lines in Fig.~\ref{fig:lattice}(a)] are zero except for the $j=3$rd site, which gives the minimized value $\hat{\bm J}_1\cdot\hat{\bm J}_3=-1$. Since any magnetic states at the $j=2$, 4, 5 and 6th sites are orthogonal to $\hat{\bm J}_1$, this state has the lowest energy for the AFM interaction $J_1<0$. As for the short-bond pairs for inter icosahedrons, i.e., $j=7$, 8, 9, 10 and 11th sites  [see solid  lines in Fig.~\ref{fig:lattice}(b)], the inner products $\hat{\bm J}_1\cdot\hat{\bm J}_j$ are zero except for the $j=11$th site, which gives the minimized value $\hat{\bm J}_1\cdot\hat{\bm J}_{11}=-1$. Since any magnetic states at the $j=7$, 8, 9 and 10th sites are orthogonal to $\hat{\bm J}_1$, this state has the lowest energy for $J_1<0$. This argument can be applied to all the other $i=2$, 3, $\cdots$, 12th sites in the similar way. Hence, it turns out that the state denoted by the light-blue-triangle symbol is realized as the ground at $\theta=\theta_0$ for the AFM interaction $J_1$ and $J_2=0$ in Fig.~\ref{fig:PD}(b). 

It is also  understandable why the state denoted by the pink-inverted triangle symbol is realized as the ground state for $J_1=0$ at $\theta=\theta_0$ in Fig.~\ref{fig:PD}(a). Focusing on the $i=1$st site [see Fig. \ref{fig:pink_inv_tri}(a)], the inner products $\hat{\bm J}_i\cdot\hat{\bm J}_j$ for the pairs with the $j=7$, 8, 9 and 10th sites [see dashed lines in Fig.~\ref{fig:lattice}(a)] for intra icosahedron are zero except for the $j=11$th site, which gives the maximized value $\hat{\bm J}_1\cdot\hat{\bm J}_{11}=+1$. Since any magnetic states at the $j=7$, 8, 9 and 10th sites are orthogonal to $\hat{\bm J}_1$, this state has the lowest energy for the FM interaction $J_2>0$. As for the inter-icosahedrons pairs with the $j=2$, 4, 5, 6 and 12th sites [see dashed lines in Fig.~\ref{fig:lattice}(b)], the inner products $\hat{\bm J}_1\cdot\hat{\bm J}_j$ are zero except for the $j=12$th site, which gives the maximized value $\hat{\bm J}_1\cdot\hat{\bm J}_{12}=+1$. Since any magnetic states at the $j=2$, 4, 5 and 6th sites are orthogonal to $\hat{\bm J}_1$, this state has the lowest energy for $J_2>0$. This argument can be applied to all the other $i=2$, $3, \cdots, 12$th sites in the similar way. Hence, it turns out that the state denoted by the pink-inverted triangle symbol is realized as the ground state at $\theta=\theta_0$ for the FM interaction $J_2$ and $J_1=0$ in Fig.~\ref{fig:PD}(a).

At $\theta=\theta_1$, it is understandable why the state denoted by the light-blue-triangle symbol is realized as the ground state for $J_1=0$. Focusing on the $i=1$st site [see Fig.~\ref{fig:lbl_tri}(b)], the inner products $\hat{\bm J}_i\cdot\hat{\bm J}_j$ for the pairs with the $j=7$, 8, 9, 10 and $11$th sites for intra icosahedron  [see dashed lines in Fig.~\ref{fig:lattice}(a)] are zero except for the $j=11$th site, which gives the maximized value $\hat{\bm J}_1\cdot\hat{\bm J}_{11}=+1$. Since any magnetic states at the $j=7$, 8, 9 and 10th sites are orthogonal to $\hat{\bm J}_1$, this state has the lowest energy for the FM interaction $J_2>0$. As for the pairs for inter icosahedrons, i.e., $j=2$, 4, 5, 6 and 12th sites [see sashed lines in Fig.~2(b)], the inner products $\hat{\bm J}_1\cdot\hat{\bm J}_j$ are zero except for the $j=12$th site, which gives the maximized value $\hat{\bm J}_1\cdot\hat{\bm J}_{12}=+1$. Since any magnetic states at the $j=2$, 4, 5 and 6th sites are orthogonal to $\hat{\bm J}_1$, this state has the lowest energy for $J_2>0$. This argument can be applied to all the other $i=2$, 3, $\cdots$, 12th sites in the similar way. Hence, it turns out that the state denoted by the light-blue-triangle symbol is realized as the ground at $\theta=\theta_1$ for the FM interaction $J_2$ and $J_1=0$ in Fig.~\ref{fig:PD}(a).

\begin{figure}[tb]
\includegraphics[width=4cm]{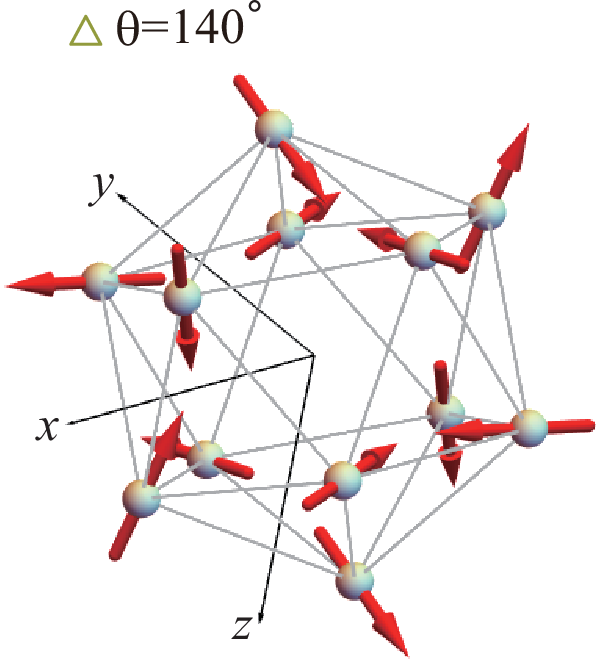}
\caption{(Color online) 
The non-collinear ferrimagnetic state for $\theta=140^{\circ}$ denoted by the green-triangle symbol in Fig.~\ref{fig:PD}. 
}
\label{fig:grn_tri}
\end{figure}

In the phase diagram shown in Fig.~\ref{fig:PD}, the ferrimagnetic order denoted by the green-triangle symbol appears in the tiny region around $\theta=140^{\circ}$ and $-J_2/J_1=50$ and around $\theta=160^{\circ}$ and $-J_2/J_1=100$. The magnetic structure is shown in Fig.~\ref{fig:grn_tri}. The magnetic space group is $R\bar{3}$ (No. 148.1.1247 in OG notation~\cite{OG} and 148.17 $R\bar{3}$ in BNS notation~\cite{BNS}), which is the same as that denoted by the light-blue triangle symbol. Figure~\ref{fig:grn_tri} is viewed from the 3-fold axis direction of rhombohedral $R\bar{3}$, i.e., the $(1,-1,-1)$ direction.

\begin{figure}[tb]
\includegraphics[width=8cm]{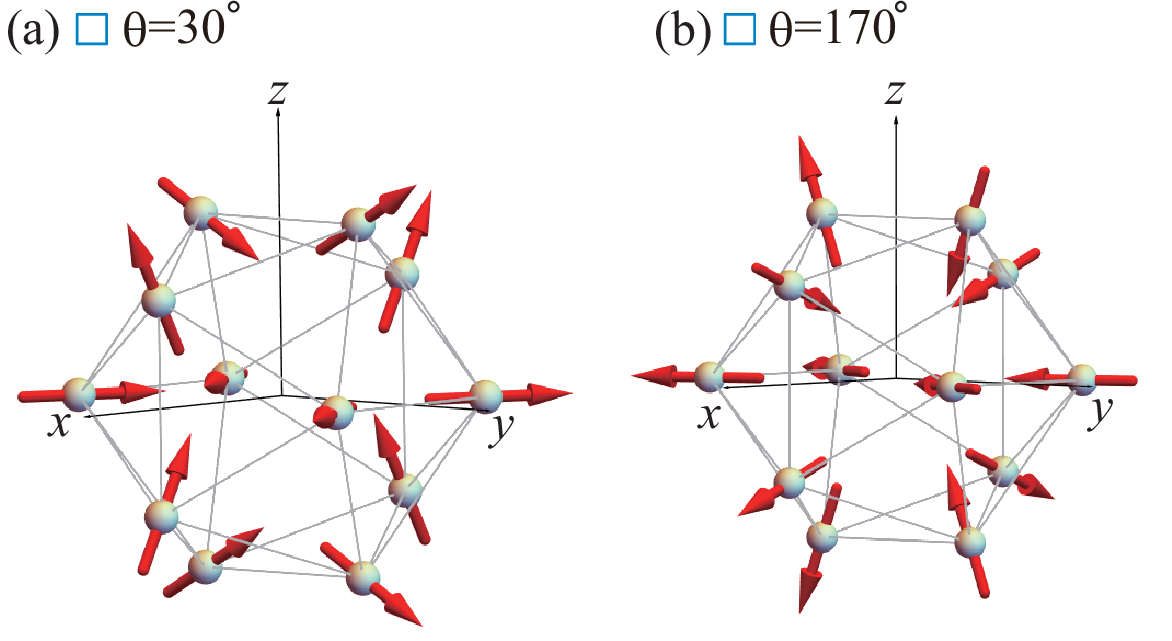}
\caption{(Color online) 
The non-collinear ferrimagnetic state for (a) $\theta=30^{\circ}$ and (b) $\theta=170^{\circ}$ denoted by the blue-square symbol in Fig.~\ref{fig:PD}. 
}
\label{fig:bl_sq}
\end{figure}

The blue-square symbol in Fig.~\ref{fig:PD} denotes the ferrimagnetic order shown in Fig.~\ref{fig:bl_sq}, where the cases for the anisotropy angle $\theta=30^{\circ}$ and $\theta=170^{\circ}$ are illustrated in Fig.~\ref{fig:bl_sq}(a) and Fig.~\ref{fig:bl_sq}(b), respectively. The magnetic space group is $C2'/m'$ (No. 12.5.70 in the OG notation~\cite{OG} and No. 12.62 $C2'/m'$ in the BNS notation~\cite{BNS}). The magnetic states shown in Figs.~\ref{fig:bl_sq}(a) and \ref{fig:bl_sq}(b) are invariant under the $180^{\circ}$ rotation around the $x$ axis with inversion of the magnetic moment (time-reversal operation). These states are also invariant under the mirror operation with respect to the mirror plane perpendicular to the $x$ axis with inversion of the magnetic moment. The total magnetic moment per icosahedron ${\bm J}_{\rm IC}$ and the total chirality ${\bm \chi}^{\rm T}$ have no component along the unique axis of the monoclinic $C2'/m'$. For instance, ${\bm J}_{\rm IC}=(0.00,5.42,3.42)$ and ${\bm \chi}^{\rm T}=(0.00,-3.41,0.19)$ for the magnetic state shown in Fig.~\ref{fig:bl_sq}(a). The degeneracy of this state is 12. 
From cubic $Im\bar{3}$ to the monoclinic $C2'/m'$, the way of choosing the unique axis among $\pm x$, $\pm y$, $\pm z$ directions is six. This six degree and time reversal one yield $6\times 2=12$. Our numerical calculation confirmed this as the number of degeneracy, 12 of the ground states, which corresponds to the number of the magnetic domains~\cite{WI2026}.
These characteristics are summarized in Table~\ref{tb:mag_state}.

\begin{figure}[tb]
\includegraphics[width=8cm]{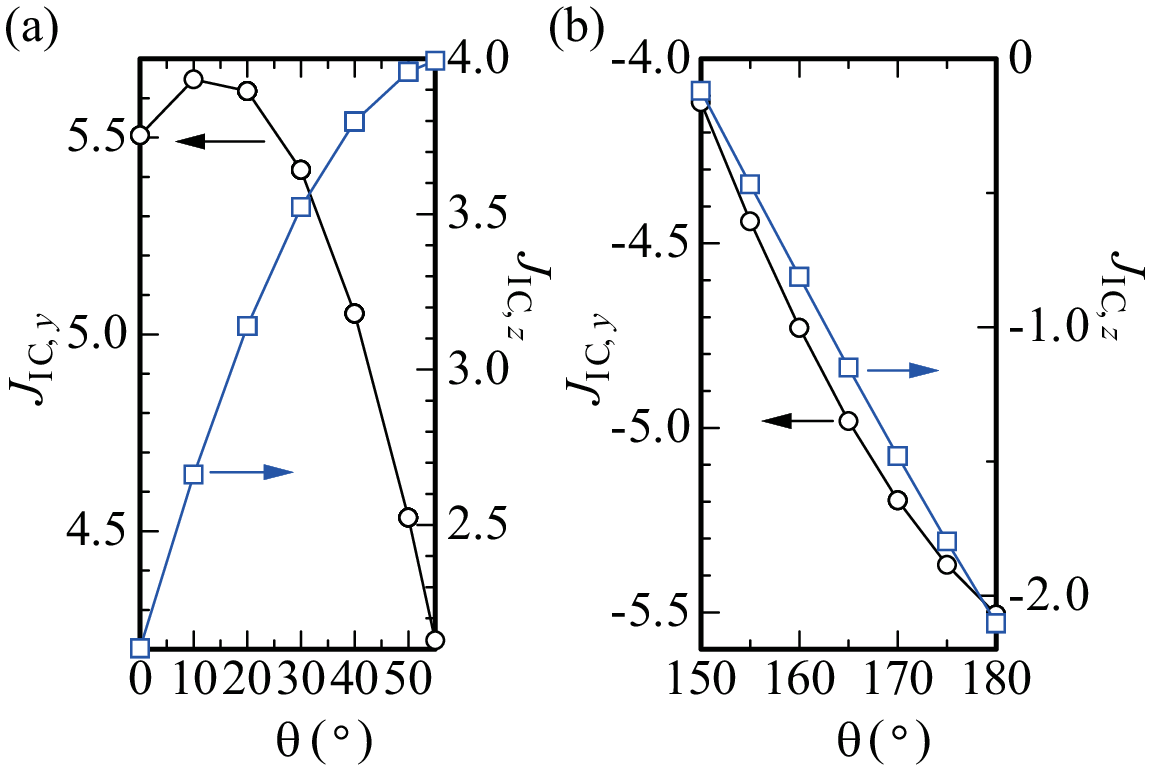}
\caption{(Color online) 
The anisotropy angle $\theta$ dependences of $J_{{\rm IC},y}$ (circle, left axis) and $J_{{\rm IC},z}$ (blue square, right axis) are shown for the ferrimagnetic order denoted by the blue-square symbol in Fig.~\ref{fig:PD} for (a) $0^{\circ}\le\theta\le 55^{\circ}$ and (b) $150^{\circ}\le\theta\le 180^{\circ}$. 
}
\label{fig:bl_sq_J}
\end{figure}

\begin{figure}[tb]
\includegraphics[width=8cm]{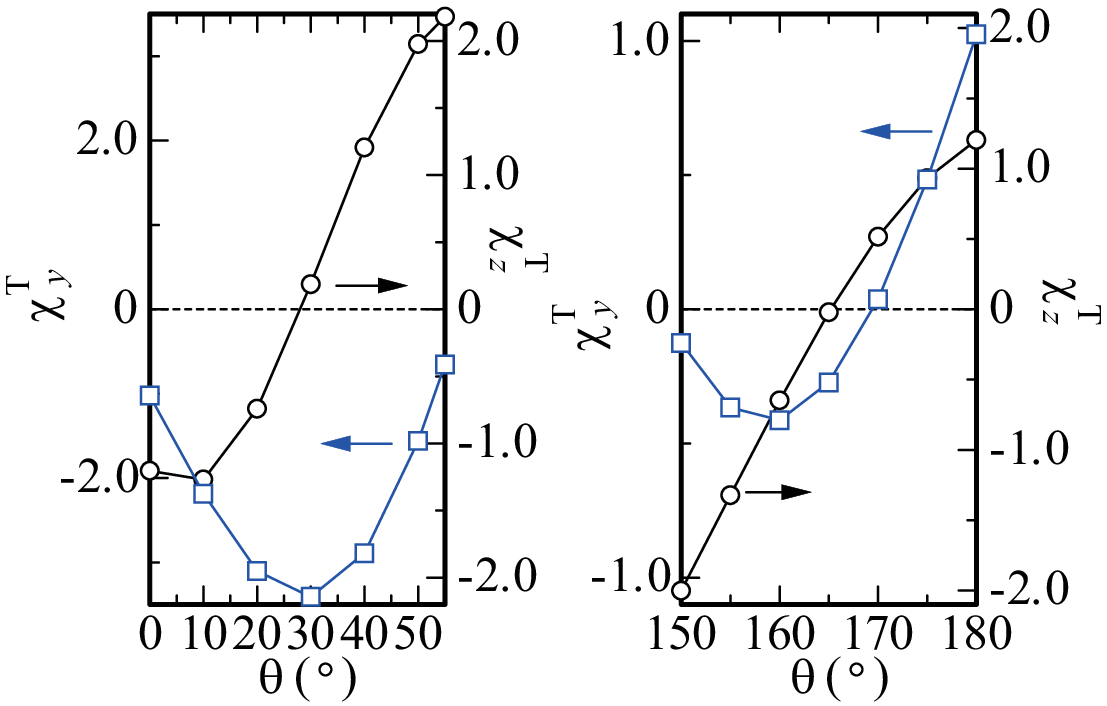}
\caption{(Color online)
The anisotropy angle $\theta$ dependence of $\chi^{\rm T}_{y}$ (blue square, left axis) and $\chi^{\rm T}_{z}$ (black circle, right axis) is shown for the ferrimagnetic order denoted by the blue-square symbol in Fig.~\ref{fig:PD} for (a) $0^{\circ}\le\theta\le 55^{\circ}$ and (b) $150^{\circ}\le\theta\le 180^{\circ}$. 
}
\label{fig:bl_sq_chiT}
\end{figure}

This ferrimagnetic order shown in Figs.~\ref{fig:bl_sq}(a) and \ref{fig:bl_sq}(b) shares the regions for $0^{\circ}\le\theta\lsim 55^{\circ}$ and for $150^{\circ}\lsim\theta\le 180^{\circ}$ in the phase diagram presented in Fig.~\ref{fig:PD}, where the $y$ component of the total magnetization per icosahedron ${\bm J}_{\rm IC}$ is zero. The anisotropy angle $\theta$ dependences of $J_{{\rm IC},y}$ and $J_{{\rm IC},z}$ are shown for both regions in Figs.~\ref{fig:bl_sq_J}(a) and \ref{fig:bl_sq_J}(b). As for the total chirality, the $x$ component of ${\bm \chi}^{\rm T}$ is zero. The anisotropy angle $\theta$ dependence of $\chi^{\rm T}_{y}$ is shown in Figs.~\ref{fig:bl_sq_chiT}(a) and \ref{fig:bl_sq_chiT}(b).

\begin{figure}[tb]
\includegraphics[width=8cm]{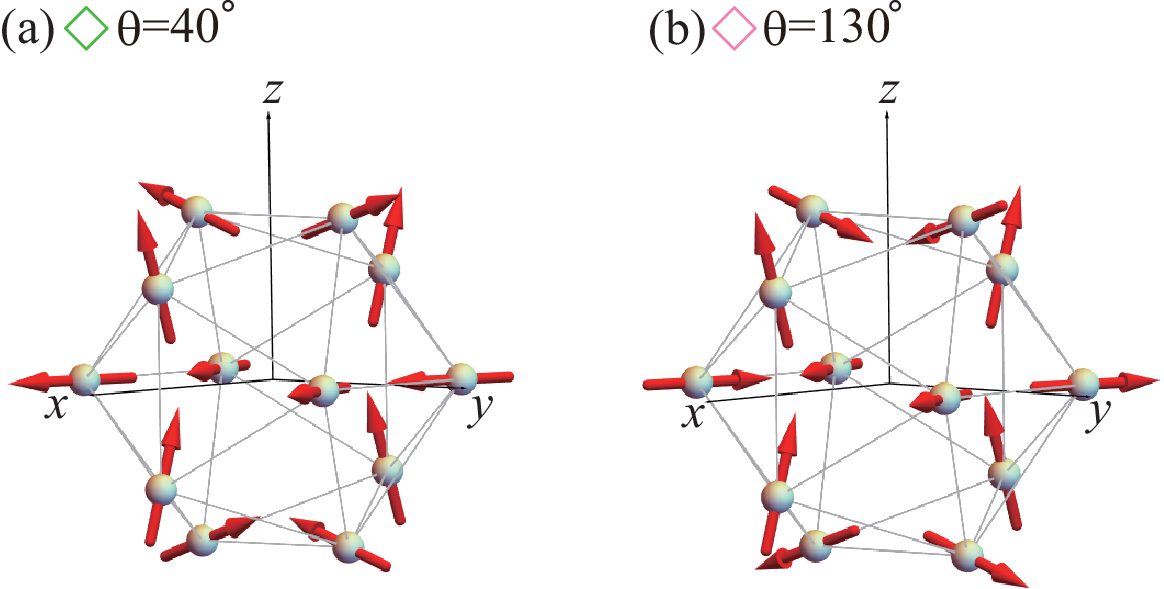}
\caption{(Color online) 
The non-collinear ferrimagnetic state for (a) $\theta=40^{\circ}$ denoted by the green-diamond symbol in Fig.~\ref{fig:PD} and (b) $\theta=130^{\circ}$ denoted by the pink-diamond symbol in Fig.~\ref{fig:PD}. 
}
\label{fig:grn_pink_dia}
\end{figure}

In the phase diagram shown in Fig.~\ref{fig:PD}, the green-diamond symbol appears in the tiny region around $\theta=40^{\circ}$ and $-J_2/J_1=50$ in the region of the ferrimagnetic order denoted by the blue-square symbol. The magnetic structure is shown in Fig.~\ref{fig:grn_pink_dia}(a). The magnetic space group is $C2'/m'$ (No. 12.5.70 in the OG notation~\cite{OG} and No. 12.62 $C2'/m'$ in the BNS notation~\cite{BNS}), which is the same as that denoted by the blue-square symbol. Namely, the magnetic state is invariant under the $180^{\circ}$ rotation around the $y$ axis with inversion of the magnetic moment and is also invariant under the mirror operation with respect to the mirror plane perpendicular to the $y$ axis with inversion of the magnetic moment.

\begin{figure}[tb]
\includegraphics[width=8cm]{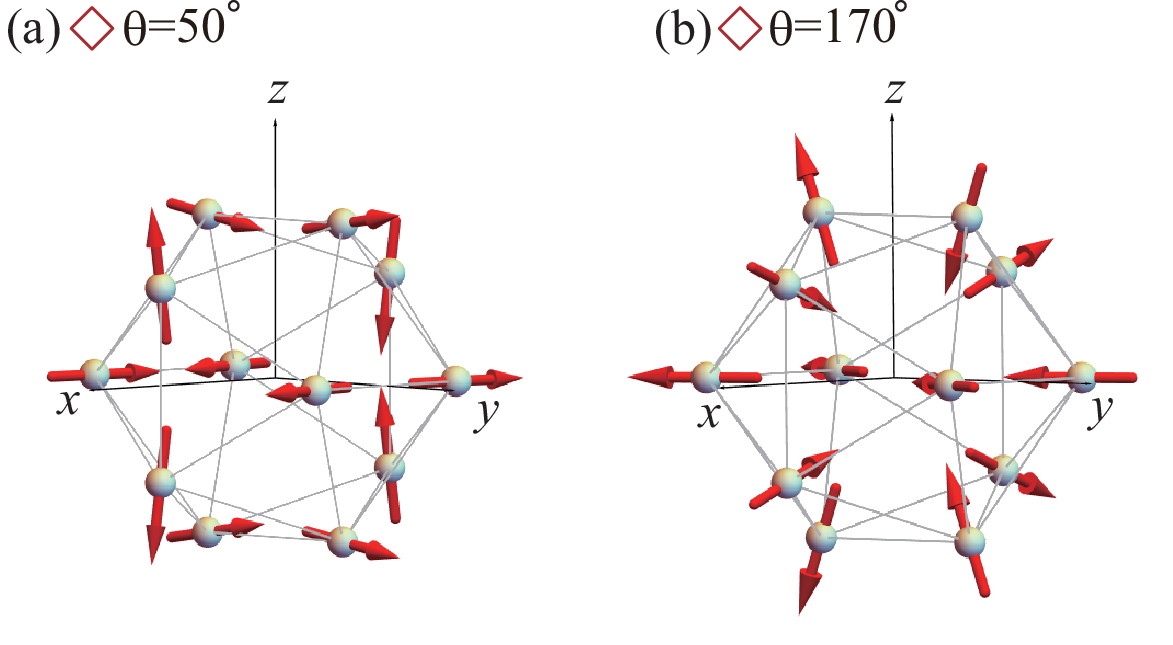}
\caption{(Color online) 
The non-collinear ferrimagnetic state for (a) $\theta=50^{\circ}$ and (b) $\theta=170^{\circ}$ denoted by the brown-diamond symbol in Fig.~\ref{fig:PD}. 
}
\label{fig:br_dia}
\end{figure}

The brown-diamond symbol in Fig.~\ref{fig:PD} denotes the ferrimagnetic order as shown in Fig.~\ref{fig:br_dia}, where the cases for the anisotropy angle $\theta=50^{\circ}$ and $\theta=170^{\circ}$ are illustrated in Fig.~\ref{fig:br_dia}(a) and Fig.~\ref{fig:br_dia}(b), respectively. The magnetic space group is $C2'/m'$ (No. 12.5.70 in the OG notation~\cite{OG} and No. 12.62 $C2'/m'$ in the BNS notation~\cite{BNS}). The magnetic states shown in Figs.~\ref{fig:br_dia}(a) and \ref{fig:br_dia}(b) are invariant under the $180^{\circ}$ rotation around the $z$ axis with inversion of the magnetic moment. These states are also invariant under the mirror operation with respect to the mirror plane perpendicular to the $z$ axis with inversion of the magnetic moment. The total magnetic moment per icosahedron ${\bm J}_{\rm IC}$ and the total chirality ${\bm \chi}^{\rm T}$ have no component along the unique axis of the monoclinic $C2'/m'$. For instance, ${\bm J}_{\rm IC}=(0.57,4.53,0.00)$ and ${\bm \chi}^{\rm T}=(3.80,-1.52,0.00)$ for the magnetic state shown in Fig.~\ref{fig:br_dia}(a). The degeneracy of this state is 12 for $-J_2/J_1\ne 2$, while the degeneracy is 24 for $-J_2/J_1=2$. 
The degeneracy corresponds to the number of magnetic domains which arises from lowering symmetry from $Im\bar{3}$ to $C2'/m'$, as noted above.
These characteristics are summarized in Table~\ref{tb:mag_state}.

\begin{figure}[tb]
\includegraphics[width=8cm]{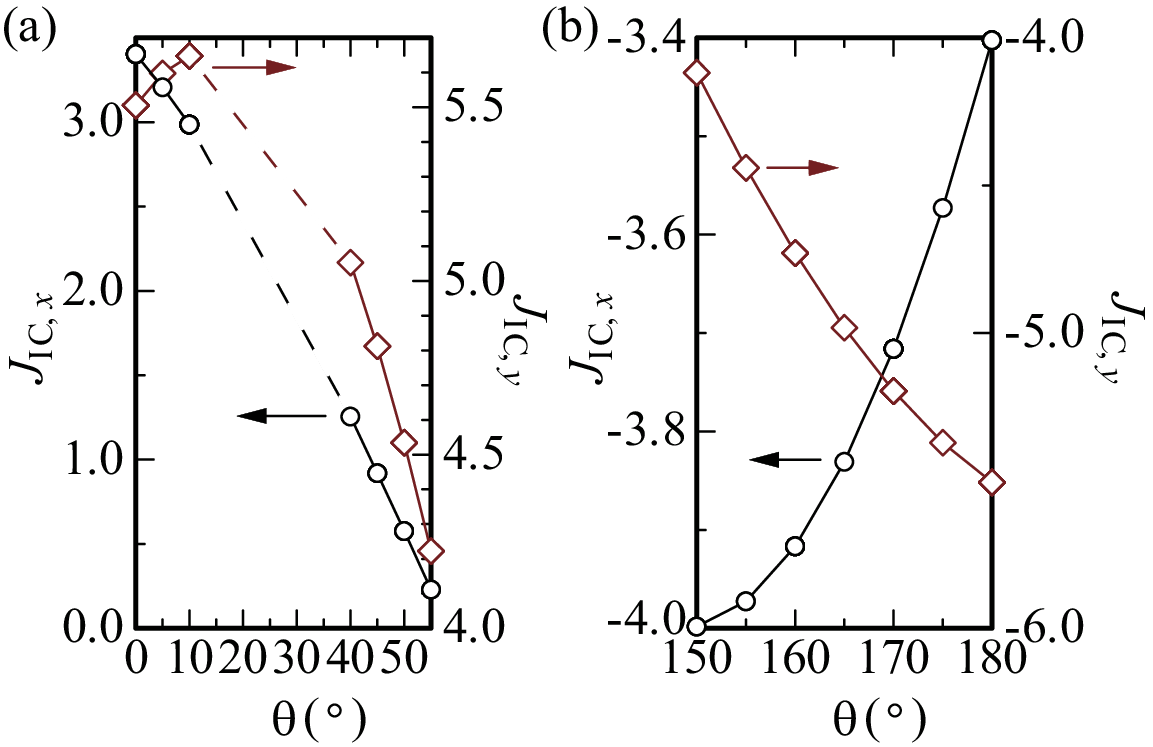}
\caption{(Color online) 
The anisotropy angle $\theta$ dependences of $J_{{\rm IC},x}$ (circle, left axis) and $J_{{\rm IC},y}$ (brown diamond, right axis) are shown for the ferrimagnetic order denoted by the brown-diamond symbol in Fig.~\ref{fig:PD} for (a) $0^{\circ}\le\theta\le 10^{\circ}$ and $40^{\circ}\le\theta\le 55^{\circ}$ and (b) $150^{\circ}\le\theta\le 180^{\circ}$. In (a), dashed lines between the ordered regimes are the guides for the eyes. 
}
\label{fig:br_dia_J}
\end{figure}

\begin{figure}[tb]
\includegraphics[width=8cm]{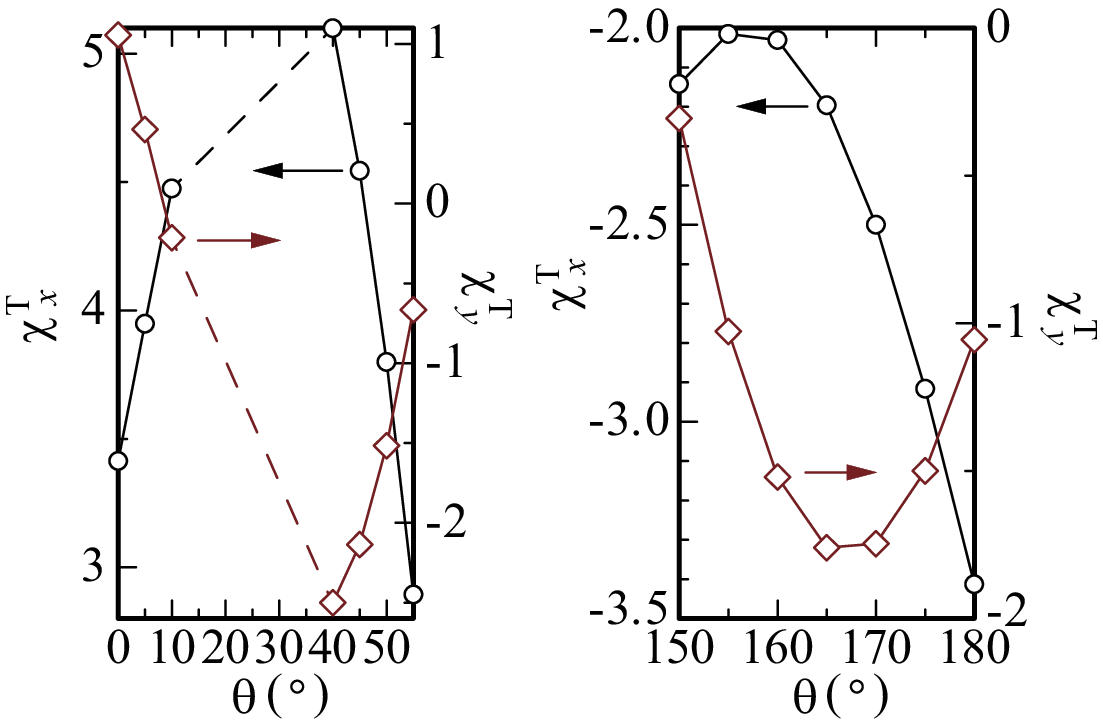}
\caption{(Color online) 
The anisotropy angle $\theta$ dependences of $\chi^{\rm T}_{x}$ (circle, left axis) and $\chi^{\rm T}_{y}$ (brown diamond, right axis) are shown for the ferrimagnetic order denoted by the brown-diamond symbol in Fig.~\ref{fig:PD} for (a) $0^{\circ}\le\theta\le 10^{\circ}$ and $40^{\circ}\le\theta\le 55^{\circ}$ and (b) $150^{\circ}\le\theta\le 180^{\circ}$. In (a), dashed lines between the ordered regimes are the guides for the eyes. 
}
\label{fig:br_dia_chiT}
\end{figure}

This ferrimagnetic order shown in Figs.~\ref{fig:br_dia}(a) and \ref{fig:br_dia}(b) shares the regions for $0^{\circ}\le\theta\lsim 10^{\circ}$, $40^{\circ}\lsim\theta\lsim 55^{\circ}$ and $150^{\circ}\lsim\theta\le 180^{\circ}$ in the phase diagram presented in Fig.~\ref{fig:PD}, where the $z$ component of the total magnetization per icosahedron ${\bm J}_{\rm IC}$ is zero. The anisotropy angle $\theta$ dependences of $J_{{\rm IC},y}$ and $J_{{\rm IC},z}$ are shown in Figs.~\ref{fig:br_dia_J}(a) and \ref{fig:br_dia_J}(b). As for the total chirality, the $z$ component of ${\bm \chi}^{\rm T}$ is zero. The anisotropy angle $\theta$ dependence of $\chi^{\rm T}_{x}$ and $\chi^{\rm T}_{y}$ are shown in Figs.~\ref{fig:br_dia_chiT}(a) and \ref{fig:br_dia_chiT}(b). 

\begin{figure}[tb]
\includegraphics[width=8cm]{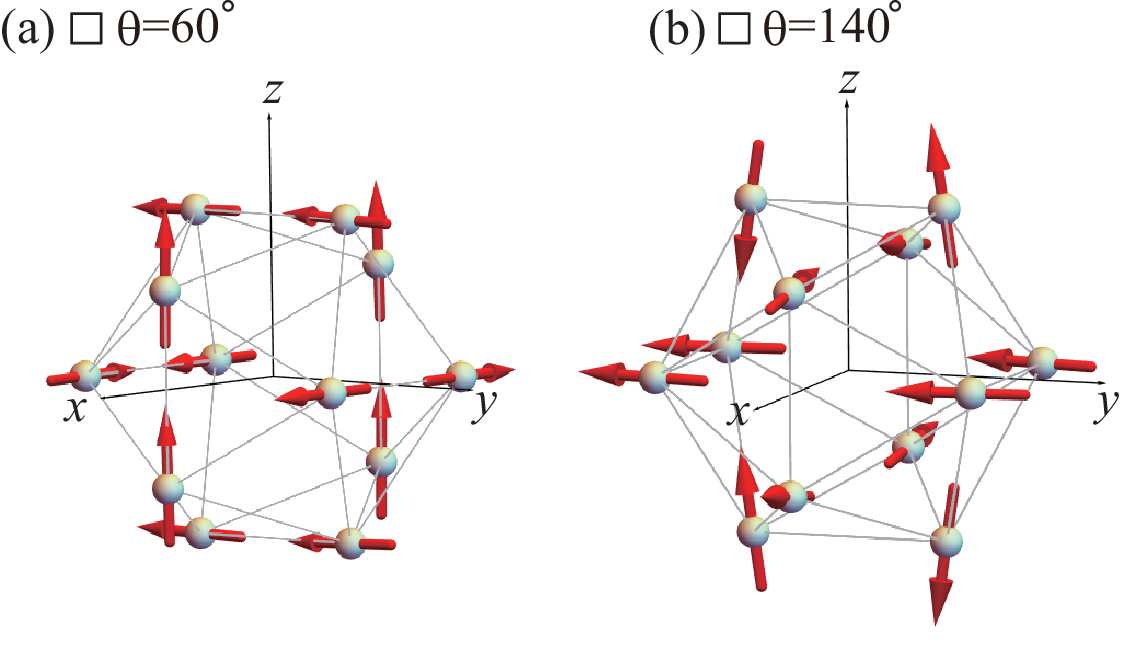}
\caption{(Color online) 
The non-collinear ferrimagnetic state for (a) $\theta=60^{\circ}$ and (b) $\theta=140^{\circ}$ denoted by the black-square symbol in Fig.~\ref{fig:PD}. 
}
\label{fig:blck_sq}
\end{figure}

The black-square symbol in Fig.~\ref{fig:PD} denotes the ferrimagnetic order as shown in Fig.~\ref{fig:blck_sq}, where the cases for the anisotropy angle $\theta=60^{\circ}$ and $\theta=140^{\circ}$ are illustrated in Fig.~\ref{fig:blck_sq}(a) and Fig.~\ref{fig:blck_sq}(b), respectively. The magnetic space group is $C2'/m'$ (No. 12.5.70 in the OG notation~\cite{OG} and No. 12.62 $C2'/m'$ in the BNS notation~\cite{BNS}). Namely, the magnetic states shown in Figs.~\ref{fig:blck_sq}(a) and (b) are invariant under the $180^{\circ}$ rotation around the $x$ axis with inversion of the magnetic moment. These states are also invariant under the mirror operation with respect to the mirror plane perpendicular to the $x$ axis with inversion of the magnetic moment. The total magnetic moment per icosahedron ${\bm J}_{\rm IC}$ and the total chirality ${\bm \chi}^{\rm T}$ have no component along the unique axis of the monoclinic $C2'/m'$. For instance, ${\bm J}_{\rm IC}=(0.00,-4.12,4.00)$ and ${\bm \chi}^{\rm T}=(0.00,-1.21,1.11)$ for the magnetic state shown in Fig.~\ref{fig:blck_sq}(a). The degeneracy of this state is 12. 
The degeneracy corresponds to the number of magnetic domains which arises from lowering symmetry from $Im\bar{3}$ to $C2'/m'$, as noted above. 
These characteristics are summarized in Table~\ref{tb:mag_state}.

\begin{figure}[tb]
\includegraphics[width=8cm]{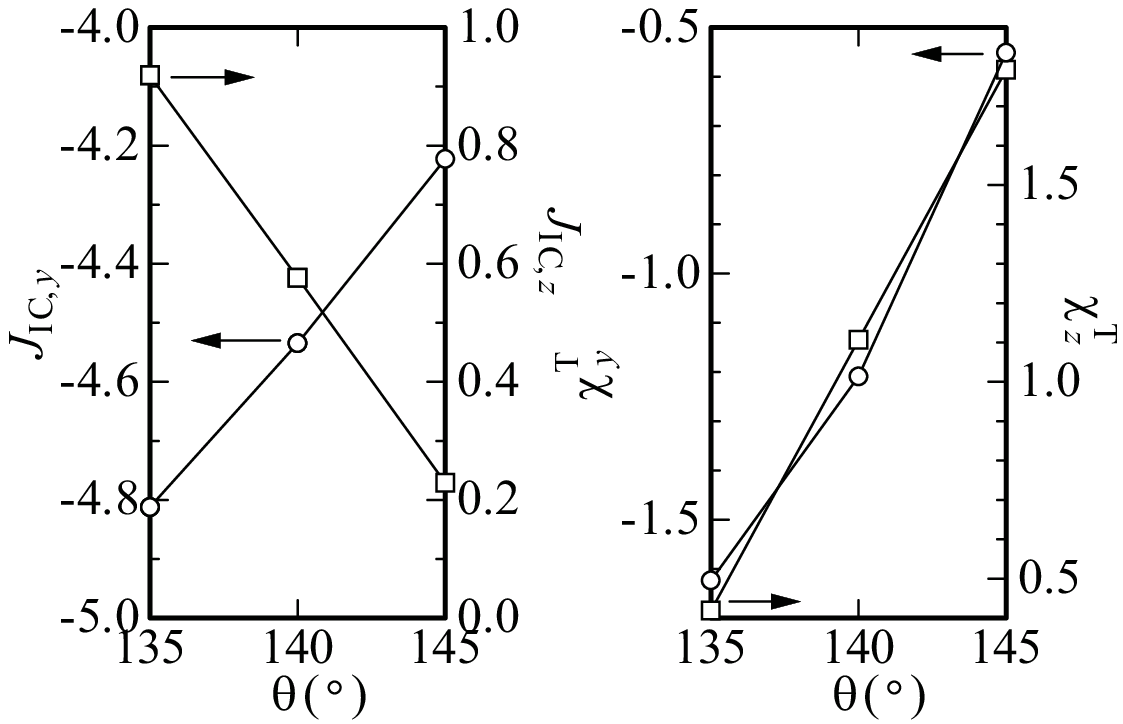}
\caption{ 
The anisotropy angle $\theta$ dependences of (a) $J_{{\rm IC},y}$ (circle, left axis) and $J_{{\rm IC},z}$ (square, right axis) and (b) $\chi^{\rm T}_{y}$ (circle, left axis) and $\chi^{\rm T}_{z}$ (square, right axis) are shown for the ferrimagnetic order denoted by the black-square symbol in Fig.~\ref{fig:PD}. 
}
\label{fig:blck_sq_J_chiT}
\end{figure}

This ferrimagnetic order shown in Figs.~\ref{fig:blck_sq}(a) and \ref{fig:blck_sq}(b) shares the regions around $\theta=60^{\circ}$ and for $135^{\circ}\lsim\theta\lsim 145^{\circ}$ in the phase diagram presented in Fig.~\ref{fig:PD}, where the $x$ components of the total magnetization per icosahedron ${\bm J}_{\rm IC}$ and the total chirality ${\bm \chi}^{\rm T}$ are zero. The anisotropy angle $\theta$ dependences of $J_{{\rm IC},y}$ and $J_{{\rm IC},z}$ are shown in Fig.~\ref{fig:blck_sq_J_chiT}(a) and those of $\chi^{\rm T}_{y}$ and $\chi^{\rm T}_{z}$ are shown in Fig.~\ref{fig:blck_sq_J_chiT}(b). 

\begin{figure}[tb]
\includegraphics[width=8cm]{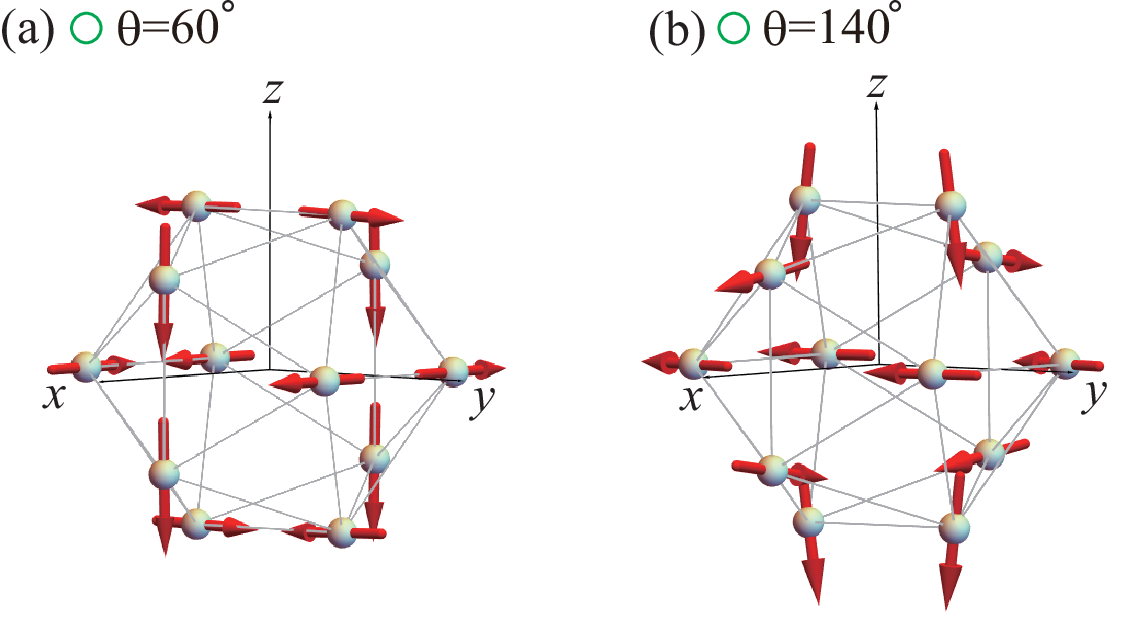}
\caption{(Color online) 
The non-collinear ferrimagnetic state for (a) $\theta=60^{\circ}$ and (b) $\theta=140^{\circ}$ denoted by the green-circle symbol in Fig.~\ref{fig:PD}. 
}
\label{fig:gr_cir}
\end{figure}

The green-circle symbol in Fig.~\ref{fig:PD} represents the ferrimagnetic order as shown in Fig.~\ref{fig:gr_cir}, where the cases for the anisotropy angle $\theta=60^{\circ}$ and $\theta=140^{\circ}$ are illustrated in Fig.~\ref{fig:gr_cir}(a) and Fig.~\ref{fig:gr_cir}(b), respectively. The magnetic space group is $C2'/m'$ (No. 12.5.70 in the OG notation~\cite{OG} and No. 12.62 $C2'/m'$ in the BNS notation~\cite{BNS}). The magnetic states shown in Figs.~\ref{fig:gr_cir}(a) and \ref{fig:gr_cir}(b) are invariant under the $180^{\circ}$ rotation around the $x$ axis with inversion of the magnetic moment. These states are also invariant under the mirror operation with respect to the mirror plane perpendicular to the $x$ axis with inversion of the magnetic moment. The total magnetic moment per icosahedron ${\bm J}_{\rm IC}$ and the total chiralitity ${\bm \chi}^{\rm T}$ have no component along the unique axis of the monoclinic $C2'/m'$. For instance, ${\bm J}_{\rm IC}=(0.00,-0.12,-4.12)$ and ${\bm \chi}^{\rm T}=(0.00,1.50,-0.40)$ for the magnetic state shown in Fig.~\ref{fig:gr_cir}(a). The degeneracy of this state is 12. 
The degeneracy corresponds to the number of magnetic domains which arises from lowering symmetry from $Im\bar{3}$ to $C2'/m'$, as noted above.  
These characteristics are summarized in Table~\ref{tb:mag_state}. 

\begin{figure}[tb]
\includegraphics[width=8cm]{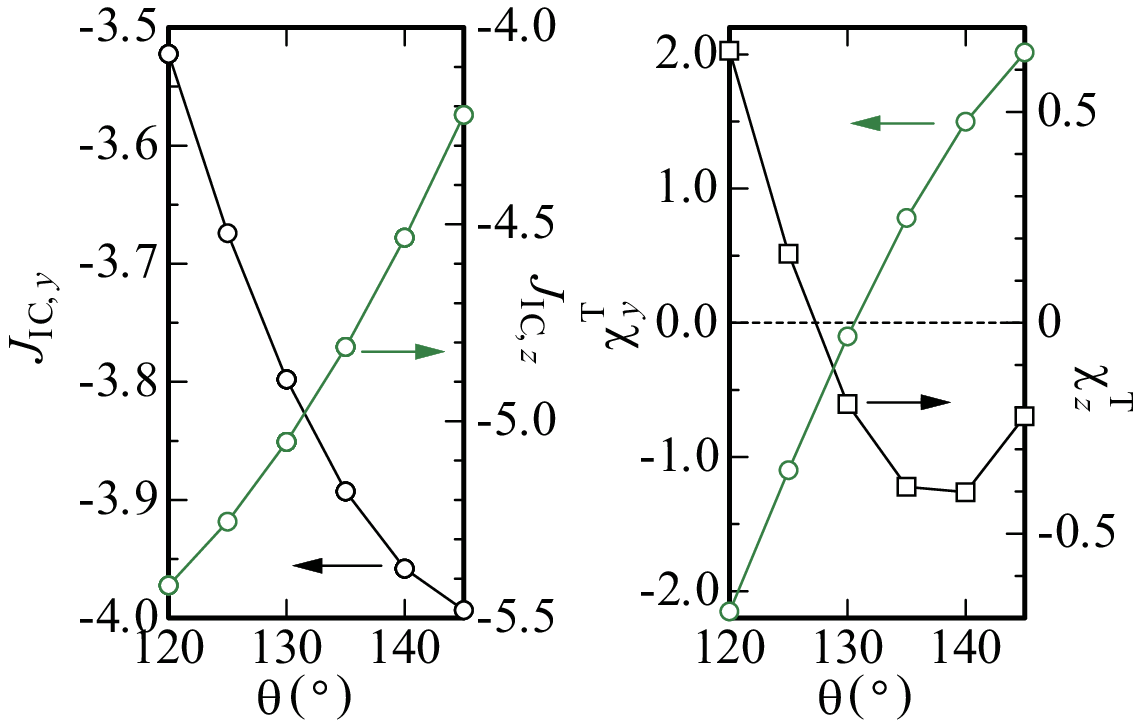}
\caption{(Color online)
The anisotropy angle $\theta$ dependences of (a) $J_{{\rm IC},y}$ (black circle, left axis) and $J_{{\rm IC},z}$ (green circle, right axis) and (b) $\chi^{\rm T}_{y}$ (green circle, left axis) and  $\chi^{\rm T}_{z}$ (black square, right axis) are shown for the ferrimagnetic order denoted by the green-circle symbol in Fig.~\ref{fig:PD}. 
}
\label{fig:gr_cir_J_chiT}
\end{figure}

This ferrimagnetic order shown in Figs.~\ref{fig:gr_cir}(a) and \ref{fig:gr_cir}(b) shares the regions around $\theta=60^{\circ}$ and for $120^{\circ}\lsim\theta\lsim 145^{\circ}$ at $-J_2/J_1\approx 2$ in the phase diagram presented in Fig.~\ref{fig:PD}, where the $x$ component of the total magnetization per icosahedron ${\bm J}_{\rm IC}$ is zero and the $x$ component of the total chirality ${\bm \chi}^{\rm T}$ is  zero. The anisotropy angle $\theta$ dependences of $J_{{\rm IC},y}$ and $J_{{\rm IC},z}$ are shown in Figs.~\ref{fig:gr_cir_J_chiT}(a) and those of $\chi^{\rm T}_{y}$ and $\chi^{\rm T}_{z}$ are shown in Fig.~\ref{fig:gr_cir_J_chiT}(b). 

In the phase diagram presented in Fig.~\ref{fig:PD}, the ferrimagnetic order denoted by the pink-diamond symbol appears in the tiny region around $\theta=130^{\circ}$ and $-J_2/J_1=100$. The magnetic structure is shown in Fig.~\ref{fig:grn_pink_dia}(b). The magnetic space group is $C2'/m'$ (No. 12.5.70 in the OG notation~\cite{OG} and No. 12.62 $C2'/m'$ in the BNS notation~\cite{BNS}), which is the same as that denoted by the green-circle symbol. Namely, the magnetic state is invariant under the $180^{\circ}$ rotation with inversion of the magnetic moment and is invariant under the mirror operation with respect to the mirror plane perpendicular to the $x$ axis with inversion of the magnetic moment. 

\newpage

\begin{table*}
\caption{
Characteristics of each magnetic order realized in the 1/1 AC. 
The symbol in the 1st column denotes the magnetic order shown in Fig.~\ref{fig:PD}. 
The 2nd to 8th columns represent 
the magnetic propagation vector ${\bm q}=\frac{2\pi}{a}(hkl)$, the magnetic space group in the OG notation, the independent magnetic position, degeneracy, the total magnetic moment vector per icosahedron ${\bm J}_{\rm IC}$, the total chirality $\chi^{\rm T}$, and the absolute value of the topological charge, respectively. 
It is noted that on the 6th and 7th columns, the states with all the components of ${\bm J}_{\rm IC}$ and ${\bm \chi}^{\rm T}$ having opposite sign are energetically degenerate. 
The lattice constant $a$ is $a=14.726~{\rm \AA}$ for cubic unit cell (i.e., the side length of a cube in Fig.~\ref{fig:lattice}). 
}
\label{tb:mag_state}
\begin{tabular}{lccllcll}
\hline
\hline
states &
$(hkl)$ & mag. & position & 
\textrm{deg.}& \textrm{${\bm J}_{\rm IC}$} & \textrm{${\bm \chi}^{\rm T}$} &
$|n|$
\\
& & sp. gr. & & & & & 
\\
\hline
black & $(000)$ & $Pm'm'2$ & $(0.187,0.306,0)$ & 12 & $(0.00,2.10,0.00)$ & $(0.00,1.12,0.00)^{\rm a}$ & $1^{\rm b}$ \\
double & & & $(0,0.187,0.306)$ & & $(0.00, 0.00, 2.10)$ & $(0.00, 0.00, 1.12)$ & 1 \\ 
square & & & $(0\textcolor{red}{.}306,0,0.187)$ & & $(0.00, 0.00,-2.10)$ & $(0.00, 0.00, -1.12)$ & 1 \\ 
Fig.~\ref{fig:blck_db_sq} & & & $(0,0.187,0.694)$ & & $(0.00,-2.10, 0.00)$ & $(0.00, -1.12, 0.00)$ & 1 \\
& & & $(0.304,0,0.813)$ & & $(2.10, 0.00, 0.00)$ & $(1.12, 0.00, 0.00)$ & 1 \\
& & & $(0.687,0.806,0.5)$ & & $(2.10, 0.00, 0.00)$ & $(1.12, 0.00, 0.00)$ & 1 \\
& & & $(0.5,0.687,0.806)$ & & & & \\
& & & $(0.806,0\textcolor{red}{.}5,0.687)$ & & & & \\
& & & $(0.5,0.687,0.194)$ & & & & \\
& & & $(0.804,0.5,0.313)$ & & & & \\
\hline
black star & $(000)$ & $Pm'm'2$ & $(0.187,0.306,0)$ & 12 & $(0.00,1.90,0.00)$ & $(0.00,0.28,0.00)^{\rm c}$ & 0 \\
Fig.~\ref{fig:blck_str} & & & $(0,0.187,0.306)$ & & $(0.00, 0.00, 1.90)$ & $(0.00, 0.00,,0.28)$ & 0 \\ 
& & & $(0,306,0,0.187)$ & & $(0.00, 1.90, 0.00)$ & $(0.00,0.28, 0.00)$ & 0 \\ 
& & & $(0,0.187,0.694)$ & & $(1.90, 0.00, 0.00)$ & $(0.28, 0.00, 0.00)$ & 0 \\
& & & $(0.304,0,0.813)$ & & $(0.00, 0.00,-1.90)$ & $(0.00, 0.00, -0.28)$ & 0 \\
& & & $(0.687,0.806,0.5)$ & & $(-1.90, 0.00, 0.00)$ & $(-0.28, 0.00, 0.00)$ & 0 \\
& & & $(0.5,0.687,0.806)$ & & & & \\
& & & $(0.806,0,5,0.687)$ & & & & \\
& & & $(0.5,0.687,0.194)$ & & & & \\
& & & $(0.804,0.5,0.313)$ & & & & \\
\hline 
pink 
& $(000)$ & $R\bar{3}$ & $(-0.170,0.103,0.265)$ & 8 & $(4.00, 4.00, 4.00)$ & $(-0.01,-0.01,-0.01)^{\rm d}$ & 0 \\
inverted & & & $(-0.427,-0.265,-0.162)$ & & $(4.00, 4.00,-4.00)$ & $(-0.01,-0.01, 0.01)$ & 0 \\
triangle & & & & & $(4.00,-4.00, 4.00)$ & $(-0.01, 0.01,-0.01)$ & 0 \\ 
Fig.~\ref{fig:pink_inv_tri}(a) & & & & & $(4.00,-4.00,-4.00)$ & $(-0.01, 0.01, 0.01)$ & 0 \\ 
\hline
light blue 
& $(000)$ & $R\bar{3}$ & $(-0.170,0.103,0.265)$ & 8 & $(-0.12,-0.12,-0.12)$ & $(-0.36, -0.36,-0.36)^{\rm e}$ & 0 \\
triangle & & & $(-0.427,-0.265,-0.162)$ & & $( 0.12,-0.12,-0.12)$ & $(0.36, -0.36, -0.36)$ & 0 \\
Fig.~\ref{fig:lbl_tri}(a) & & & & & $(-0.12,-0.12, 0.12)$ & $(-0.36, -0.36,0.36)$ & 0 \\
& & & & & $(0.12,-0.12, 0.12)$ & $(0.36, -0.36, 0.36)$ & 0 \\
\hline 
blue 
& $(000)$ & $C2'/m'$ & $(0.187,0.306,0.187)$ &12 & $(0.00, 5.42, 3.52)$ & $(0.00, -3.41, 0.19)^{\rm f}$ & 0 \\
square & & & $(0,0.187,0.306)$ & & $(0.00, 5.42,-3.52)$ & $(0.00, -3.41,-0.19)$ & 0 \\
Fig.~\ref{fig:bl_sq}(a)& & & $(0.306,0,0.119)$ & & $(5.42, 3.52, 0.00)$ & $(-3.41, 0.19, 0.00)$ & 0 \\ 
& & & $(0.306,0,0.493)$ & & $(3.52, 0.00, 5.42)$ & $(0.19, 0.00, -3.41)$ & 0 \\
& & & & & $(5.42,-3.52, 0.00)$ & $(-3.41,-0.19, 0.00)$ & 0 \\
& & & & & $(3.52, 0.00,-5.42)$ & $(0.19, 0.00, 3.41)$ & 0 \\ 
\hline
brown & $(000)$ & $C2'/m'$ & $(0.187,0.306,0.187)$ & 12 & $(0.58, 4.53, 0.00)$ & $(3.80, -1.52, 0.00)^{\rm g}$ & 0 \\
diamond & & & $(0,0.187,0.306)$ & & $(0.00, 0.58, 4.53)$ & $(0.00, 3.80, -1.52)$ & 0 \\
Fig.~\ref{fig:br_dia}(a) & & & $(0.306,0,0.119)$ & & $(-0.58, 4.53, 0.00)$ & $(-3.80, -1.52, 0.00)$ & 0 \\ 
& & & $(0.306,0,0.493)$ & & $(0.00, 0.58,-4.53)$ & $( 0.00, 3.80, 1.52)$ & 0 \\ 
& & & & & $(4.53, 0.00, 0.58)$ & $( -1.52, 0.00, 3.80)$ & 0 \\ 
& & & & & $(4.53, 0.00,-0.58)$ & $(-1.52, 0.00,-3.80)$ & 0 \\ 
\hline 
black 
& $(000)$ & $C2'/m'$ & $(0.187,0.306,0.187)$ & 12 & $(0.00,-4.53, 0.58)$ & $( 0.00,-1.21, 1.11)^{\rm h}$ & 0 \\
square & & & $(0,0.187,0.306)$ & & $(0.00,-4.53,-0.58)$ & $( 0.00,-1.21, -1.11)$ & 0 \\
Fig.~\ref{fig:blck_sq}(a)& & & $(0.306,0,0.119)$ & & $(0.58, 0.00,-4.53)$ & $(1.11, 0.00,-1.21)$ & 0 \\
& & & $(0.306,0,0.493)$ & & $(4.53, 0.58, 0.00)$ & $(1.21, 1.11, 0.00)$ & 0 \\
& & & & & $(0.58, 0.00, 4.53)$ & $(1.11, 0.00, 1.21)$ & 0 \\
& & & & & $(4.53,-0.57, 0.00)$ & $(1.21,-1.11, 0.00)$ & 0 \\
\hline
green 
& $(000)$ & $C2'/m'$ & $(0.187,0.306,0.187)$ & 12 & $(0.00,-3.96,-4.53)$ & $( 0.00, 1.50, -0.40)^{\rm i}$ & 0 \\
circle & & & $(0,0.187,0.306)$ & & $(-3.96,-4.53, 0.00)$ & $( 1.50, -0.40, 0.00)$ & 0 \\
Fig.~\ref{fig:gr_cir}(a)& & & $(0\textcolor{red}{.}306,0,0.119)$ & &$(0.00,-3.96, 4.53)$ & $( 0.00, 1.50, 0.40)$ & 0 \\
& & & $(0.306,0,0.493)$ & & $(3.96,-4.53, 0.00)$ & $(-1.50, -0.40, 0.00)$ & 0 \\
& & & & & $(4.53, 0.00,-3.95)$ & $( 0.40, 0.00, 1.50)$ & 0 \\
& & & & & $(4.53, 0.00, 3.95)$ & $( 0.40, 0.00, -1.50)$ & 0 \\
\hline
orange square 
\\
Fig.~\ref{fig:or_sq} & $(111)$ & $I_{\rm P}m'\bar{3}'$ & $(0.187,0.306,0)$ & 2 & {\bf 0} & {\bf 0} & $3^{\rm j}$ 
\\
\hline
\end{tabular}
\\
$^{\rm a}$ \footnotesize{$\theta=0^{\circ}$ case. For the $\theta$ dependence, see Fig.~\ref{fig:blck_db_sq_J_chiT}.} 
$^{\rm b}$ \footnotesize{$\theta<\theta_{{\rm c}0}^{\circ}$ case. }
$^{\rm c}$ \footnotesize{$\theta=30^{\circ}$ case. For the $\theta$ dependence, see Fig.~\ref{fig:blck_str_J_chiT}.} 
$^{\rm d}$ \footnotesize{$\theta=60^{\circ}$ case. For the $\theta$ dependence, see Fig.~\ref{fig:pink_inv_tri_J_chiT}.} 
$^{\rm e}$ \footnotesize{$\theta=60^{\circ}$ case. For the $\theta$ dependence, see Fig.~\ref{fig:lbl_tri_J_chiT}.} 
$^{\rm f}$ \footnotesize{$\theta=30^{\circ}$ case. For the $\theta$ dependence, see Fig.~\ref{fig:bl_sq_J} and Fig.~\ref{fig:bl_sq_chiT}.} 
$^{\rm g}$ \footnotesize{$\theta=50^{\circ}$ case. For the $\theta$ dependence, see Fig.~\ref{fig:br_dia_J} and Fig.~\ref{fig:br_dia_chiT}.} 
$^{\rm h}$ \footnotesize{$\theta=140^{\circ}$ case. For the $\theta$ dependence, see Fig.~\ref{fig:blck_sq_J_chiT}}. 
$^{\rm i}$ \footnotesize{$\theta=140^{\circ}$ case. For the $\theta$ dependence, see Fig.~\ref{fig:gr_cir_J_chiT}.} 
$^{\rm j}$ \footnotesize{$\theta_{{\rm c}1}<\theta<\theta_{{\rm c}2}$ case.}
\end{table*}

\newpage

\subsection{Antiferromagnetic order}

\begin{figure}[tb]
\includegraphics[width=6.5cm]{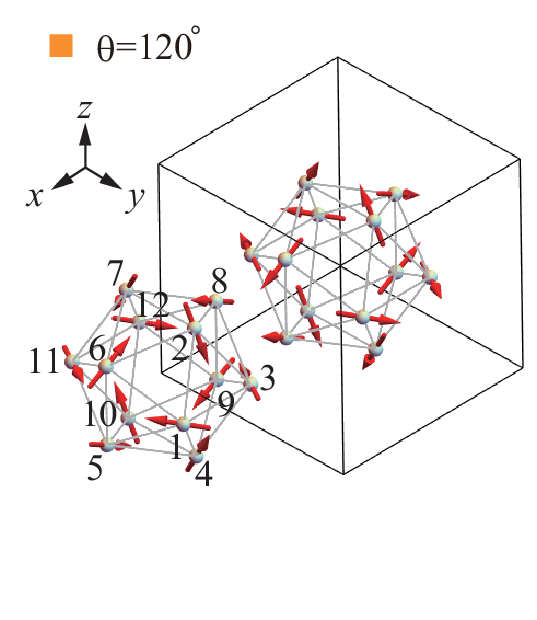}
\caption{(Color online) 
The whirling-anti-whirling state for $\theta=120^{\circ}$ denoted by the filled-orange-square symbol in Fig.~\ref{fig:PD}. 
The number labels the rare-earth site on the icosahedron. 
}
\label{fig:or_sq}
\end{figure}

The antiferromagnetic order is formed in the region denoted by the filled-orange-square symbol in the phase diagram presented in Fig.~\ref{fig:PD}. 
The magnetic moments swirl in the view from the 3-fold-axis direction as seen in Fig.~\ref{fig:or_sq} where the case for the anisotropy angle $\theta=120^{\circ}$ is shown. By inverting all the magnetic moments at 12 vertices of the corner icosahedron in the bcc unit cell, the magnetic moments at 12 vertices of the center icosahedron are obtained. The topological charge of the magnetic state at the corner icosahedron is $n=+3$ and that at the center icosahedron is $n=-3$. 

As shown in Fig.~\ref{fig:PD}, the whirling-anti-whirling state is realized around $\theta=90^{\circ}$ for whole region of $0\le -J_2/J_1\le\infty$. This is because all the inner products $\hat{\bm J}_i\cdot\hat{\bm J}_j$ becomes negative for the N.N. bonds and positive for the N.N.N. bonds around $\theta=90^{\circ}$, which are stabilized by the AFM N.N. interaction $J_1<0$ and the FM N.N.N. interaction $J_2>0$. The stabilization mechanism is explained in Appendix.

It is noted that this antiferromagnetic order was experimentally observed in the 1/1 AC Au-Al-Tb~\cite{Sato2019,Nawa2023,Labib2023}, which is referred to as the whirling-anti-whirling state. 
The magnetic space group of the whirling-anti-whirling state is $I_{\rm P}m'\bar{3}'$~\cite{Nawa2023,Labib2023,WI2026}. 
The total magnetic moment per icosahedron of the whirling state and the anti-whirling state is zero ${\bm J}_{\rm IC}={\bf 0}$. The total chirality of the whirling state and the anti-whirling state is zero ${\bm \chi}^{\rm T}={\bf 0}$. 
Our numerical calculation has shown that the degeneracy of the ground state is 2 indicating that the magnetic state where all the magnetic moments are inverted is energetically degenerate. These characteristics are summarized in Table~\ref{tb:mag_state}. 

It is noted that there exist the characteristic anisotropy angles $\theta_{{\rm c}i}$ $(i=1,2,3)$, which are defined as 
$\theta_{{\rm c}1}=79.475\dots^{\circ}$, $\theta_{{\rm c}2}=103.570\dots^{\circ}$, and $\theta_{{\rm c}3}=127.665\dots^{\circ}$~\cite{W2026}. 
The topological charges of the magnetic texture at the corner and center icosahedron shown in Fig.~\ref{fig:or_sq} are $(+1,-1)$ respectively for $\theta<\theta_{{\rm c}1}$ and $(-3,+3)$ for $\theta_{{\rm c}1}<\theta<\theta_{{\rm c}2}$, while the former and the latter are $(+3,-3)$ respectively for $\theta_{{\rm c}2}<\theta<\theta_{{\rm c}3}$ and $(-1,+1)$ respectively for $\theta>\theta_{{\rm c}3}$. In the phase diagram presented in Fig.~\ref{fig:PD}, the absolute values of the topological charge are plotted. 

In ref.~\citen{WPNAS}, it was theoretically shown that the field-induced topological as well as metamagnetic phase transition takes place in the whirling-anti-whirling order. Furthermore, it was pointed out theoretically that the topological Hall effect is expected to appear in the electrical conductivity $\sigma_{yz}^{\rm T}$ and $\sigma_{xy}^{\rm T}$ after the metamagnetic transition under the magnetic field applied along the $z$-axis (2-fold axis) direction~\cite{WPNAS}. Recently, the magnetic-field-direction dependence has been studied theoretically for the whirling-anti-whirling order~\cite{WICM2024} and it has been shown that in case that the magnetic-field direction is continuously changed from the $(001)$ to $(111)$ direction, the degeneracy of the ground state is lifted so that the single domain is achieved.

\section{Discussion}
\label{sec:disc}

\subsection{Characteristic anisotropy angle}

On the basis of Eq.~(\ref{eq:tp}), we have calculated the topological charge for each magnetic texture on icosahedron. 
The magnetic textures on the corner icosahedron and the center icosahedron shown in Fig.~\ref{fig:blck_db_sq} are characterized by the topological charge $n=-1$ and $n=+1$, respectively. In the phase diagram shown in Fig.~\ref{fig:PD}, this ferrimagnetic order denoted by the black double-square symbol is characterized by $n=0$ for $\theta\gsim 20^{\circ}$. In this section, we analyze the topological charge of these states as a function of the anisotropy angle $\theta$, which results in the characteristic anisotropy angle $\theta_{{\rm c}0}$. 

\begin{figure}[tb]
\includegraphics[width=8cm]{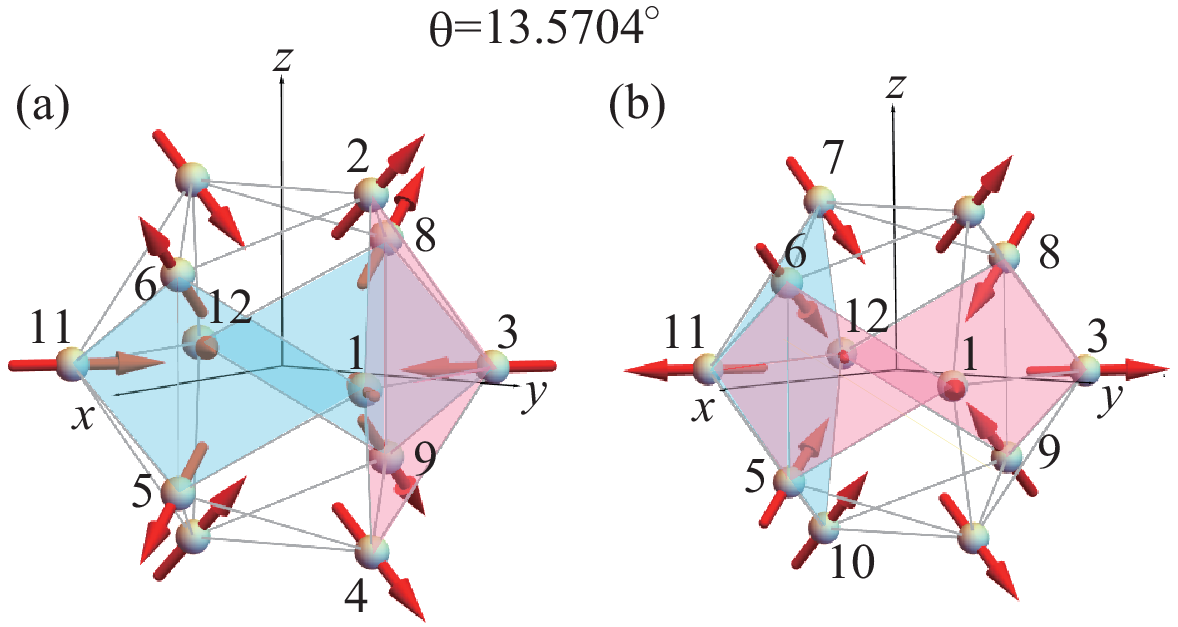}
\caption{(Color online) 
The magnetic states for $\theta=13.5704^{\circ}$ of (a) the corner icosahedron and (b) the center icosahedron in the ferrimagnetic order denoted by the black-double-square symbol in the phase diagram 
shown in 
Fig.~\ref{fig:PD}. The number labels the rare-earth site on the icosahedron. 
}
\label{fig:blck_db_sq_2IC}
\end{figure}

When the magnetic-moment vectors at the three vertices of the triangular surface on the icosahedron, $i, j$ and $k$ sites, satisfy the relation
\begin{eqnarray}
{\bm J}_i\cdot{\bm J}_j+{\bm J}_j\cdot{\bm J}_k+{\bm J}_k\cdot{\bm J}_i=-1,
\label{eq:Jinpsum3}
\end{eqnarray}
the denominator of Eq.~(\ref{eq:SI_tri}) becomes zero. For the magnetic state denoted by the black double-square symbol, we have identified that at $\theta=\theta_{{\rm c}0}\equiv 13.5704\dots^{\circ}$ Eq.~(\ref{eq:Jinpsum3}) is satisfied. 
Namely, as for the magnetic state on the corner icosahedron shown in Fig.~\ref{fig:blck_db_sq}, for $\theta=\theta_{{\rm c}0}^{-}\equiv\theta_{{\rm c}0}-\delta$ with $\delta$ being a positive definite, $\frac{\Omega_{ijk}}{4\pi}=\frac{1}{4}$ for the $(i,j,k)=(3,9,8)$, $(8,9,12)$, $(1,6,5)$, and $(5,6,11)$ vertices and $\frac{\Omega_{ijk}}{4\pi}=-\frac{1}{4}$ for the $(i,j,k)=(7,12,11)$ and $(10,11,12)$ vertices, whose triangular surfaces are colored in the light pink and the light blue, respectively, in Fig.~\ref{fig:blck_db_sq_2IC}(b). As $\theta$ increases from $\theta=\theta_{{\rm c}0}^{-}$ to $\theta=\theta_{{\rm c}0}^{+}\equiv\theta_{{\rm c}0}+\delta$, $\frac{\Omega_{ijk}}{4\pi}$ changes from $\frac{1}{4}$ $(-\frac{1}{4})$ to $-\frac{1}{4}$ $(\frac{1}{4})$ for the $ijk$ triangular surfaces colored in the light pink (the light blue). For the other triangular surfaces on the icosahedron, $\frac{\Omega_{ijk}}{4\pi}$ changes continuously. Namely, the change in the topological charge at $\theta=\theta_{{\rm c}0}$ is $(\frac{1}{4}-(-\frac{1}{4}))\times 2+(-\frac{1}{4}-\frac{1}{4})\times 4=-1$. Hence, the topological charge changes from $n=+1$ to $n=+1-1=0$ as $\theta$ changes from $\theta_{{\rm c}0}^{-}$ to $\theta_{{\rm c}0}^{+}$.

As for the magnetic state on the center icosahedron shown in Fig.~\ref{fig:blck_db_sq}, for $\theta=\theta_{{\rm c}0}^{-}$, $\frac{\Omega_{ijk}}{4\pi}=\frac{1}{4}$ for the $(i,j,k)=(3,9,8)$, $(8,9,12)$, $(1,6,5)$ and $(5,6,11)$ vertices and $\frac{\Omega_{ijk}}{4\pi}=-\frac{1}{4}$ for the $(i,j,k)=(7,12,11)$ and $(10,11,12)$ vertices, whose triangular surfaces are colored in the light pink and the light blue, respectively, in Fig.~\ref{fig:blck_db_sq_2IC}(b). For the other triangular surfaces on the icosahedron, $\frac{\Omega_{ijk}}{4\pi}$ changes continuously. Hence, as $\theta$ increases from $\theta=\theta_{{\rm c}0}^{-}$ to $\theta=\theta_{{\rm c}0}^{+}$, $\frac{\Omega_{ijk}}{4\pi}$ changes from $\frac{1}{4}$ $(-\frac{1}{4})$ to $-\frac{1}{4}$ $(\frac{1}{4})$. Namely, the change in the topological charge at $\theta=\theta_{{\rm c}0}$ is $(\frac{1}{4}-(-\frac{1}{4}))\times 2+(-\frac{1}{4}-\frac{1}{4})\times 4=-1$. Hence, the topological charge changes from $n=-1$ to $n=-1+1=0$ as $\theta$ changes from $\theta_{{\rm c}0}^{-}$ to $\theta_{{\rm c}0}^{+}$. 

Hence, each magnetic texture on the corner and center icosahedrons for the ferrimagnetic order denoted by the black double-square symbol for $\theta\ge 20^{\circ}$ in Fig.~\ref{fig:PD} is characterized by $n=0$. 

As for the antiferromagnetic order denoted by the filled orange-square symbol in Fig.~\ref{fig:PD}, there exist characteristic anisotropy angles, which satisfy Eq.~(\ref{eq:Jinpsum3})~\cite{W2026}. 
Similarly to the ferrimagnetic order denoted by the black-double-square symbol, the characteristic anisotropy angles $\theta_{{\rm c}i}$ $(i=1, 2, 3)$ were identified for the whirling-anti-whirling order~\cite{W2026}. 
Those are $\theta_{{\rm c}1}=79.475\dots^{\circ}$, $\theta_{{\rm c}2}=103.570\dots^{\circ}$, and $\theta_{{\rm c}3}=127.665\dots^{\circ}$. For detail of the derivation of $\theta_{{\rm c}1}$, $\theta_{{\rm c}2}$, and $\theta_{{\rm c}3}$, readers refer to ref.~\citen{W2026}. 

In the ground-state phase diagram presented in Fig.~\ref{fig:PD}, the characteristic anisotropy angles $\theta_{{\rm c}i}$ $(i=0, 1, 2, 3)$ are plotted as the vertical dotted lines. 

\subsection{Topological Hall effect}

As presented in Table~\ref{tb:mag_state}, the total chirality ${\bm \chi}^{\rm T}$ emerges in every ferrimagnetic orders. 
Since ${\bm \chi}^{\rm T}$ plays the role as an emergent fictitious magnetic field, the topological Hall effect is expected to appear via the relation of Eq.~(\ref{eq:sigma}). 
In the ferrimagnetic orders denoted by the black-double-square symbol and the black-star symbol, 
${\bm \chi}^{\rm T}$ emerges along the 2-fold axis. Hence, the topological Hall effect is expected to appear in the perpendicular components of the electrical conductivity $\sigma_{\mu\nu}^{\rm T}$. 

As for the ferrimagnetic orders denoted by the pink-inverted triangle and the light-blue triangle, ${\bm \chi}^{\rm T}$ emerges along the 3-fold axis of the rhombohedral $R\bar{3}$. Hence, the topological Hall effect is expected to appear in the perpendicular components of the electrical conductivity $\sigma_{\mu\nu}^{\rm T}$. For instance, in the ferrimagnetic order shown in Fig.~\ref{fig:lbl_tri}(a), the total chirality emerges as ${\bm \chi}^{\rm T}\propto (111)$. Hence, the topological Hall conductivity 
is  
expected to appear in $\sigma_{xy}^{\rm T}$, $\sigma_{yz}^{\rm T}$, and $\sigma_{zx}^{\rm T}$.
In the ferrimagnetic orders denoted by the brown-diamond symbol and the black-square symbol, the total chirality ${\bm \chi}^{\rm T}$ has no component along the unique axis of the monoclinic $C2'/m'$. For instance, in the case shown in Fig.~\ref{fig:br_dia}(a), the unique axis is the $z$ axis. 
Hence, the topological Hall effect is expected to appear in $\sigma_{zx}^{\rm T}$ and  $\sigma_{yz}^{\rm T}$.

\section{Summary}
\label{sec:summary}

By performing complete enumeration of all configurations of magnetic moments, 
we have determined the ground-state phase diagram of the effective model for the AFM N.N. interaction $J_1$ and the FM N.N.N. interaction $J_2$ on the 1/1 AC. 
Depending on the values of $J_2/J_1$ and the magnetic-anisotropy angle $\theta$, various non-collinear magnetic structures are stabilized. Around $\theta=90^{\circ}$, antiferromagnetic order appears, while the ferrimagnetic orders appear for the other $\theta$ near $\theta=0^{\circ}$ and $180^{\circ}$. We have identified the magnetic space group as $I_{\rm P}m'\bar{3}'$ for the antiferromagnetic order and $R\bar{3}$, $C2'/m'$ and $Pm'm'2$ for the ferrimagnetic orders. 

We have shown that the whirling-anti-whirling state is stabilized around the anisotropy angle $\theta=90^{\circ}$ for $J_1=0$ and also for $J_2=0$ because all inner products ${\bm J}_i\cdot{\bm J}_j$ are negative for the N.N. pairs and positive for the N.N.N. pairs for intra and inter icosahedrons. We have also shown analytically that at $\theta=\theta_0=58.28^{\circ}$ and $\theta=\theta_1=148.28^{\circ}$, noncollinear ferrimagnetic state becomes the ground state. These provide solid ground for understanding the stabilization mechanism of noncollinear AFM and FM orders  in the phase diagram of $\theta$ and $-J_2/J_1$. 

We have revealed the topological as well as magnetic property of each ground state with degeneracy, which is useful to analyze experimental observations of the magnetic structures in the rare-earth-based 1/1 approximants with consideration of the effects of the magnetic domains. So far, the ferrimagnetic order denoted by the pink-inverted triangle and the antiferromagnetic order denoted by the filled-orange-square symbol in our phase diagram presented in Fig.~\ref{fig:PD} were experimentally observed. Experimental exploration of the other type of non-collinear magnetic orders which have been predicted in our study in the rare-earth-based 1/1 ACs 
is  
 interesting future subject. 

\begin{acknowledgment}
The author acknowledges K. Momma for providing the software VESTA to draw the magnetic moments with kind instructions~\cite{Momma}. 
This work was supported by JSPS KAKENHI Grant Numbers JP22H04597, JP22H01170, JP23K17672, and JP24H01675.
\end{acknowledgment}

\appendix
\section{Stabilization Mechanism of the Whirling-Anti-Whirling State}
\label{sec:WAW}

In this Appendix, we analyze the reason why the whirling-anti-whirling state is stabilized around $\theta=90^{\circ}$ for whole range of $0\le -J_2/J_1\le\infty$ in the ground-state phase diagram shown in Fig.~\ref{fig:PD}.

\begin{figure}[tb]
\includegraphics[width=6cm]{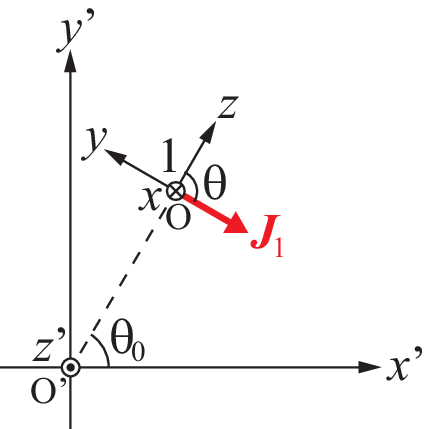}
\caption{(Color online)
The local coordinate $xyz$ set at the $i=1$st site on the icosahedron in the global coordinate $x'y'z'$ whose origin O' is set at the center of the icosahedron. The magnetic easy axis lies in the mirror plane, i.e., $yz$ plane  [see Fig.~\ref{fig:Tsai_cluster}(f)]. The direction of the magnetic moment ${\bm J}_1$ is defined by the anisotropy angle $\theta$ from the pseudo 5-fold axis, i.e., $z$ axis in the mirror plane. 
}
\label{fig:lc_gb_co}
\end{figure}

Let us focus on the 1st site in Fig.~\ref{fig:or_sq} where the whirling-anti-whirling state is realized. 
The magnetic moment at the $i=1$st site for the whiling state is given by ${\bm J}_1=(v_{x'},v_{y'},0)$, which is illustrated in the local $xyz$ coordinate [see Fig.~\ref{fig:Tsai_cluster}(f)] and global $x'y'z'$ coordinate in Fig.~\ref{fig:lc_gb_co}. When the anisotropy angle $\theta$ is around $90^{\circ}$, the following conditions are satisfied as 
\begin{eqnarray}
v_{x'}>0, v_{y'}<0,
\label{eq:vx_vy_1}
\\
|v_{x'}|>|v_{y'}|.
\label{eq:vx_vy_2}
\end{eqnarray}
The magnetic moments at the N.N. sites for intra icosahedron [see solid lines in Fig.~\ref{fig:lattice}(a)] are given by ${\bm J}_2=(0,v_{x'},v_{y'})$, ${\bm J}_3=(-v_{x'},v_{y'},0)$, ${\bm J}_4=(0, v_{x'}, -v_{y'})$, ${\bm J}_5=(v_{y'}, 0, -v_{x'})$, and ${\bm J}_6=(v_{y'},0,v_{x'})$.  Then, the inner products ${\bm J}_1\cdot{\bm J}_j$ yield 
\begin{eqnarray}
{\bm J}_1\cdot{\bm J}_j=v_{x'}v_{y'}<0
\label{eq:JJ} 
\end{eqnarray}
for the $j=2$, 4, 5, and 6th site and 
\begin{eqnarray}
{\bm J}_1\cdot{\bm J}_3=-v_{x'}^2+v_{y'}^2<0.
\label{eq:JJ2}
\end{eqnarray} 
Here, the conditions of Eq.~(\ref{eq:vx_vy_1}) and (\ref{eq:vx_vy_2}) are used to evaluate the sign of Eq.~(\ref{eq:JJ}) and Eq.~(\ref{eq:JJ2}), respectively.

 The magnetic moments at the N.N. sites for inter icosahedrons where the anti-whirling state is realized [see solid lines in Fig.~\ref{fig:lattice}(b)] are given by ${\bm J}_7=(0,v_{x'},-v_{y'})$, ${\bm J}_8=(v_{y'},0,-v_{x'})$, ${\bm J}_9=(v_{y'},0,v_{x'})$, ${\bm J}_{10}=(0,v_{x'},v_{y'})$, and ${\bm J}_{11}=(-v_{x'},v_{y'},0)$. Then, the inner products ${\bm J}_1\cdot{\bm J}_j$ yield 
 \begin{eqnarray}
 {\bm J}_1\cdot{\bm J}_j=v_{x'}v_{y'}<0
 \label{eq:JJ3}
 \end{eqnarray} 
 for the $j=7$, 8, 9, and 10th site and 
 \begin{eqnarray}
 {\bm J}_1\cdot{\bm J}_{11}=-v_{x'}^2+v_{y'}^2<0. 
\label{eq:JJ4}
 \end{eqnarray}
Here, the conditions of Eq.~(\ref{eq:vx_vy_1}) and (\ref{eq:vx_vy_2}) are used to evaluate the sign of Eq.~(\ref{eq:JJ3}) and Eq.~(\ref{eq:JJ4}), respectively.

These results indicate that all the inner products ${\bm J}_1\cdot{\bm J}_j<0$ with the $j$th site in the N.N. pairs for intra icosahedron and inter icosahedron. We confirmed that ${\bm J}_i\cdot{\bm J}_j<0$ holds for the other $i=2$, 3, $\cdots$, 12th site with the $j$th site in the N.N. pairs for intra icosahedron and inter icosahedron. This implies that for the AFM N.N. interaction $J_1<0$, the whirling-anti-whirling state is strongly stabilized at least in case that the N.N.N. interaction is zero $J_2=0$.

The condition of Eq.~(\ref{eq:vx_vy_1}) is evaluated as $\theta_0<\theta<\theta_0+\pi/2$. From the coordinate of the $i=1$st site shown in Fig.~\ref{fig:lc_gb_co}, $\theta_0$ is estimated as $\theta_0=58.28^{\circ}$ (Here, we consider the case that icosahedron is regular icosahedron, where the position of the $i=1$st site is given by $(1,\tau,0)$ and then  $\theta_0=\tan^{-1}\tau$ holds, as an ideal case). Then, the above condition leads to $58.28^{\circ}<\theta<148.28^{\circ}$, which turns out to be expressed as 
\begin{eqnarray}
\theta_0<\theta<\theta_1.
\label{eq:cond1}
\end{eqnarray}

The condition of Eq.~(\ref{eq:vx_vy_2}) is evaluated as $\theta_0<\theta<\theta_0+\pi/4$, which leads to $58.28^{\circ}<\theta<103.28^{\circ}$, which turns out to be expressed as 
\begin{eqnarray}
\theta_0<\theta<\theta_{{\rm c}2}. 
\label{eq:cond2}
\end{eqnarray}

When the condition Eq.~(\ref{eq:cond2}) is satisfied, Eq.~(\ref{eq:cond1}) is also satisfied. This gives a qualitative explanation for why the whirling-anti-whirling state is stabilized around $\theta=90^{\circ}$ for $J_2=0$ in Fig.~\ref{fig:PD}(b).

Next, let us analyze the N.N.N. pairs. Focusing on the 1st site in Fig. 25, the inner products with the $j$th site in the N.N.N. pairs for intra icosahedron where the whirling state is realized [see dashed lines in Fig.~\ref{fig:lattice}(a)] yield
\begin{eqnarray}
{\bm J}_1\cdot{\bm J}_j=-v_{x '}v_{y '}>0
\label{eq:JJ_NNN1}
\end{eqnarray}
for the $j=7$, 8, 9, and 10th site and 
\begin{eqnarray}
{\bm J}_1\cdot{\bm J}_{11}=v_{ x'}^2-v_{ y'}^2>0.
\label{eq:JJ_NNN2}
\end{eqnarray}
Here, the conditions of Eq.~(\ref{eq:vx_vy_1}) and Eq.~(\ref{eq:vx_vy_2}) are used to evaluate the sign of Eq.~(\ref{eq:JJ_NNN1}) and Eq.~(\ref{eq:JJ_NNN2}), respectively.

The inner products with the $j$th site in the N.N.N. pairs for inter icosahedron where the anti-whirling state is realized [see dashed lines in Fig.~\ref{fig:lattice}(b)] yield
\begin{eqnarray}
{\bm J}_1\cdot{\bm J}_j=-v_{x'}v_{y'}>0
\label{eq:JJ_NNN3}
\end{eqnarray}
for the $j=2$, 4, 5, and 6th site and 
\begin{eqnarray}
{\bm J}_1\cdot{\bm J}_{12}=v_{ x'}^2+v_{ y'}^2>0.
\label{eq:JJ_NNN4}
\end{eqnarray}
Here, the conditions of Eq.~(\ref{eq:vx_vy_1}) is used to evaluate the sign of Eq.~(\ref{eq:JJ_NNN3}).

These results indicate that all the inner products ${\bm J}_1\cdot{\bm J}_j>0$ with the $j$th site in the N.N.N. pairs for intra and inter icosahedron. We confirmed that ${\bm J}_i\cdot{\bm J}_j>0$ holds for the other $i=2$, 3, $\cdots$, 12th site with the $j$th site in the N.N. pairs for intra and inter icosahedron. This implies that for the FM N.N.N. interaction $J_2>0$, the whirling-anti-whirling state is strongly stabilized at least in case that the N.N. interaction is zero $J_1=0$.

Equations~(\ref{eq:JJ_NNN1}) and (\ref{eq:JJ_NNN3}) are satisfied when $\theta$ is in the range of Eq. (\ref{eq:cond1}). Equation~(\ref{eq:JJ_NNN2}) is satisfied when $\theta$ is in the range of Eq.~(\ref{eq:cond2}). When the condition Eq.~(\ref{eq:cond2}) is satisfied, Eq.~(\ref{eq:cond1}) is also satisfied. 
This gives a qualitative explanation for why the whirling-anti-whirling state is stabilized around $\theta=90^{\circ}$ for $J_1=0$ in Fig.~\ref{fig:PD}(a).

Since in both the limits of $J_2=0$ and $J_1=0$ in the phase diagram (see Fig.~\ref{fig:PD}), the whirling-anti-whirling state is stabilized around $\theta=90^{\circ}$, even in the finite region of $-J_2/J_1$, i.e., $0<-J_2/J_1<\infty$, the phase survives. This is understandable as starting from both limits, the effect of perturbation of $J_2>0$ or $J_1<0$ provides no frustration but just gives additional contribution to further stabilize the whirling-anti-whirling state.


* swata@mns.kyutech.ac.jp


\begin{thebibliography}{99}
\bibitem{Shechtman} D. Shechtman, I. Blech, D. Gratias, and J. W. Cahn, Metallic phase with long-range orientational order and no translational symmetry, {Phys. Rev. Lett.} {\bf 53}, 1951 (1984). 
\bibitem{Tsai} A. P. Tsai, J. Q. Guo, E. Abe, H. Takakura, and T. J. Sato, A stable binary quasicrystal, {Nature} {\bf 408}, 537 (2000). 
\bibitem{Takakura} H. Takakura, C. P. G\'{o}mez, A. Yamamoto, M. de. Boissieu, and A. P. Tsai, Atomic structure of the binary icosahedral Yb-Cd quasicrystal, {Nat. Mater.} {\bf 6}, 58 (2007).
\bibitem{Fisher1999} I. R. Fisher, K. O. Cheon, A. F. Panchula, P. C. Canfield, M. Chernikov, H. R. Ott, and K. Dennis, Magnetic and transport properties of single-grain $R$-Mg-Zn icosahedral quasicrystals [{\it R}=Y, (Y$_{1-x}$Gd$_x$), (Y$_{1-x}$Tb$_x$) Tb, Dy, Ho, and Er], Phys. Rev. B {\bf 59}, 308 (1999). 
\bibitem{Sato2000} T. J. Sato, H. Takakura, A. P. Tsai, K. Shibata, K. Ohoyama, and K. H. Andersen, Antiferromagnetic spin correlations in the Zn-Mg-Ho icosahedral quasicrystal,
Phys. Rev. B {\bf 61}, 476 (2000). 
\bibitem{Goldman2013} A. I. Goldman, T. Kong, A. Kreyssig, A. Jesche, M. Ramazanoglu, K. W. Dennis, S. L. Bud'ko and P. C. Canfield., A family of binary magnetic icosahedral quasicrystals based on rare earths and cadmium, Nat. Mater. {\bf 12}, 714 (2013). 
\bibitem{Das2023} P. Das, A. Kreyssig, G. S. Tucker, A. Podlesnyak, F. Ye, M. Matsuda, T. Kong, S. L. Bud'ko, P. C. Canfield, R. Flint, P. P. Orth, T. Yamada, R. J. McQueeney, and A. I. Goldman, Short-range magnetic correlations in quasicrystalline $i$-Tb-Cd, 
Phys. Rev. B {\bf 108}, 134421 (2023). 
\bibitem{Tamura2010} R. Tamura, Y. Muro, T. Hiroto, K. Nishimoto and T. Takabatake, Long-range magnetic order in the quasicrystalline approximant Cd$_6$Tb, Phys. Rev. B {\bf 82}, 220201(R) (2010). 
\bibitem{Mori} A. Mori, H. Ota, S. Yoshiuchi, K. Iwakawa, Y. Taga, Y. Hirose, T. Takeuchi, E. Yamamoto, Y. Haga, F. Honda, R. Settai and Y. $\bar{\rm O}$nuki, Electrical and magnetic properties of quasicrystal approximants RCd$_6$ (R: Rare Earth), J. Phys. Soc. Jpn. {\bf 81}, 024720 (2012). 
\bibitem{Das} P. Das, P.-F. Lory, R. Flint, T. Kong, T. Hiroto, S. L. Bud'ko, P. C. Canfield, M. de Boissieu, A. Kreyssig, and A. I. Goldman, Crystal electric field excitations in the quasicrystal approximant TbCd$_6$ studied by inelastic neutron scattering, Phys. Rev. B {\bf 95}, 054408 (2017).
\bibitem{Jazbec} S. Jazbec, S. Kashimoto, P. Ko$\check{\rm z}$elj, S. Vrtnik, M. Jagodi$\check{\rm c}$, Z. Jagli$\check{\rm c}$i$\acute{\rm c}$, and J. Dolin$\check{\rm s}$ek, Schottky effect in the $i$-Zn-Ag-Sc-Tm icosahedral quasicrystal and its 1/1 Zn-Sc-Tm approximant, Phys. Rev. B {\bf 93}, 054208 (2016). 
\bibitem{Ishikawa2018} A. Ishikawa, T. Fujii, T. Takeuchi, T. Yamada, Y. Matsushita, and R. Tamura, Antiferromagnetic order is possible in ternary quasicrystal approximants, Phys. Rev. B {\bf 98}, 220403(R) (2018).
\bibitem{Sato2019} T. J. Sato, A. Ishikawa, A. Sakurai, M. Hattori, M. Avdeev, and R. Tamura, Whirling spin order in the quasicrystal approximant Au$_{72}$Al$_{14}$Tb$_{14}$, Phys. Rev. B {\bf 100}, 054417 (2019).
\bibitem{Nawa2023} K. Nawa, M. Avdeev, A. Ishikawa, H. Takakura, C.-W. Wang, D. Okuyama, R. Murasaki, R. Tamura, and T. J. Sato, Magnetic properties of the quasicrystal approximant Au$_{65}$Ga$_{21}$Tb$_{14}$, Phys. Rev. Mater. {\bf 7}, 054412 (2023).
\bibitem{Labib2023} F. Labib, K. Nawa, S. Suzuki, H.-C. Wu, A. Ishikawa, K. Inagaki, T. Fujii, K. Kinjo, T. J. Sato, and R. Tamura, 
Unveiling exotic magnetic phase diagram of a non-Heisenberg quasicrystal approximant, 
Mater. Today Phys. {\bf 40}, 101321 (2023). 
\bibitem{Labib_ACAF_2025} F. Labib, T. Sugimoto, H. Takakura, A. Ishikawa, and R. Tamura, Identification of two distinct antiferromagnetic phases in the Au-Al-Gd quasicrystal approximant, Phys. Rev. B {\bf 111}, 174409 (2025). 
\bibitem{Hiroto2013} T. Hiroto, G. H. Gebresenbut, C. P. G\'omez, Y. Muro, M. Isobe, Y. Ueda, K. Tokiwa and R. Tamura, Ferromagnetism and re-entrant spin-glass transition in quasicrystal approximants Au-SM-Gd (SM = Si, Ge), J. Phys.: Condens. Matter {\bf 25}, 216004 (2013). 
\bibitem{Hiroto2014} T. Hiroto, K. Tokiwa and R. Tamura, Sign of canted ferromagnetism in the quasicrystal approximants Au-SM-R (SM = Si, Ge and Sn/R=Tb, Dy and Ho), J. Phys.: Condens. Matter {\bf 26}, 216004 (2014). 
\bibitem{Ishikawa2016} A. Ishikawa, T. Hiroto, K. Tokiwa, T. Fujii, and R. Tamura, Composition-driven spin glass to ferromagnetic transition in the quasicrystal approximant Au-Al-Gd, Phys. Rev. B {\bf 93}, 024416 (2016). 
\bibitem{Shiino2022} T. Shiino, G. H. Gebresenbut, C. P. G\'omez, U. H\"{a}ussermann, P. Nordblad, A. Rydh, and R. Mathieu, Examination of the critical behavior and magnetocaloric effect of the ferromagnetic Gd-Au-Si quasicrystal approximants, Phys. Rev. B {\bf 106}, 174405 (2022). 
\bibitem{Hiroto} T. Hiroto, T. J. Sato, H. Cao, T. Hawai, T. Yokoo, S. Itoh and R. Tamura, Noncoplanar ferrimagnetism and local crystalline-electric-field anisotropy in the quasicrystal approximant Au$_{70}$Si$_{17}$Tb$_{13}$, J. Phys.: Condens. Matter {\bf 32}, 415802 (2020). 
\bibitem{Gebresenbut} G. H. Gebresenbut, T. Shiino, M. S. Andersson, N. Qureshi, O. Fabelo, P. Beran, D. Qvarngard, P. Henelius, A. Rydh, R. Mathieu, P. Nordblad, and C. P. G\'omez, Effect of pseudo-Tsai cluster incorporation on the magnetic structures of $R$-Au-Si (R=Tb,Ho) quasicrystal approximants, Phys. Rev. B {\bf 106}, 184413 (2022). 
\bibitem{Labib2025} F. Labib, H. Takakura, A. Ishikawa, T. Fujii, and R. Tamura, Derivation of a Nonstoichiometric 1/1 Quasicrystal Approximant from a Stoichiometric 2/1 Quasicrystal Approximant and Maximization of the Magnetocaloric Effect, J. Am. Chem Soc. {\bf 147}, 33424 (2025). 
\bibitem{Imura} K. Imura, K. Nobe, K. Deguchi, M. Matsunami, H. Miyazaki, A. Yasui, E. Ikenaga and N. K. Sato, First observation of heavy fermion behavior in Ce-based icosahedral approximant, J. Phys. Soc. Jpn. {\bf 86}, 093702 (2017).
\bibitem{Sawano} T. Sawano, T. Shiino, K. Imura, K. Deguchi, S. Ohhashi, A.-P. Tsai, and N. K. Sato, Local quantum fluctuations in Kondo quasicrystal approximant Ag-In-(Ce$_x$Y$_{1-x}$), J. Phys. Soc. Jpn. {\bf 89}, 014703 (2020).
\bibitem{Muro} Y. Muro, T. Fukuhara, T. Namiki, T. Kuwai, A. Sakurai, A. Ishikawa, S. Suzuki and R. Tamura, Composition effect of Kondo behavior in Au-Al-Ce quasicrystalline approximants, Mater. Trans. {\bf 62}, 321 (2021). 
\bibitem{Labib2024} F. Labib, N. Fujita, H. Takakura, S. Ohhashi, T. Shiino, A.-P. Tsai, and R. Tamura, Formation and structure of 1/1 cubic approximants in the Cd-Mg-Ce System: The identification of a new structural variant, J. Alloy. Compd. {\bf 954}, 170151 (2024).
%
\bibitem{Suzuki2021} S. Suzuki, A. Ishikawa, T. Yamada, T. Sugimoto, A. Sakurai, and R. Tamura, Magnetism of Tsai-type quasicrystal approximants, Mater. Trans. {\bf 62}, 298 (2021). 
\bibitem{Tamura2025} R. Tamura, T. Abe, S. Yoshida, Y. Shimozaki, S. Suzuki, A. Ishikawa, F. Labib, M. Avdeev, K. Kinjo, K. Nawa, and T. J. Sato, Observation of antiferromagnetic order in a quasicrystal, Nat. Phys. {\bf 21}, 974 (2025). 
\bibitem{Tamura2021} R. Tamura, A. Ishikawa, S. Suzuki, T. Kotajima, Y. Tanaka, T. Seki, N. Shibata, T. Yamada, T. Fujii, C.-W. Wang, M. Avdeev, K. Nawa, D. Okuyama, and T. J. Sato., Experimental observation of long-range magnetic order in icosahedral quasicrystals, J. Am. Chem. Soc. {\bf 143(47)}, 19938 (2021).
\bibitem{Takeuchi2023} R. Takeuchi, F. Labib, T. Tsugawa, Y. Akai, A. Ishikawa, S. Suzuki, T. Fujii, and R. Tamura, 
High phase-purity and composition-tunable ferromagnetic icosahedral quasicrystal, {Phys. Rev. Lett.} {\bf 130}, 176701 (2023).
\bibitem{NKSato2022} N. K. Sato, T. Ishimasa, K. Deguchi, and K. Imura, Effects of electron correlation and geometrical frustration on magnetism of icosahedral quasicrystals and approximants -- An attempt to bridge the gap between quasicrystals and heavy fermions, J. Phys. Soc. Jpn. {\bf 91}, 072001 (2022).
%
%
%
\bibitem{Sugimoto2016} T. Sugimoto, T. Tohyama, T. Hiroto, and R. Tamura, Phenomenological magnetic model in Tsai-type approximants, J. Phys. Soc. Jpn. {\bf 85}, 053701 (2016). 
\bibitem{Miyazaki2020} H. Miyazaki, T. Sugimoto, K. Morita, and T. Tohyama, Magnetic orders induced by RKKY interaction in Tsai-type quasicrystalline approximant Au-Al-Gd, Phys. Rev. Mater. {\bf 4}, 024417 (2020). 
\bibitem{Sugimoto2024} T. Sugimoto, S. Suzuki, R. Tamura, and T. Tohyama, Revisiting the Magnetic Orders in Heisenberg Model for Tsai-Type Quasicrystal Approximant, J. Phys. Soc. Jpn. {\bf 93}, 045001 (2024). 
%
\bibitem{WYTF2025} S. Watanabe, T. Yamada, H. Takakura, and N. Fujita, Monte Carlo study on critical exponents of the classical Heisenberg model in ferromagnetic icosahedral quasicrystal, Phys. Rev. Research {\bf 7}, 043113 (2025). 
%
\bibitem{WK2021} S. Watanabe and M. Kawamoto, Crystalline electronic field in rare-earth based quasicrystal and approximant: Analysis of quantum critical Au-Al-Yb quasicrystal and approximant J. Phys. Soc. Jpn. {\bf 90}, 063701 (2021). 
\bibitem{WPNAS} S. Watanabe, Topological magnetic textures and long-range orders in Terbium-based quasicrystal and approximant, Proc. Natl. Acad. Sci. USA. {\bf 118(43)}, e2112202118 (2021).
\bibitem{WI2023} S. Watanabe and T. Iwasaki, Crystalline electronic field and magnetic anisotropy in Dy-based icosahedral quasicrystal and approximant, Phys. Rev. B {\bf 108}, 045110 (2023). 
%
\bibitem{OG} W. Opechowski and R. Guccione, {\it Magnetism}, edited by G. T. Rado and H. Suhl, Vol. 2A (Academic Press, New York, 1965), Chap. 3. P.105.
\bibitem{BNS} N. V. Belov, N. N. Neronova, and T. S. Smirnova, Kristallografiya {\bf 2}, 315 (1957); Soviet Phys.-Cryst {\bf 2}, 311 (1957). 
%
\bibitem{WFK2026} S. Watanabe, S. Furuya, and M. Kawamura, 
Odd-parity crystalline electric field and multipoles in icosahedral quasicrystal and approximant: Magnetic toroidal and electric dipoles in Ce-based systems, 
submitted. 
\bibitem{WI2026} S. Watanabe and T. Iwasaki, Non-collinear and non-coplanar magnetic orders in 1/1 periodic approximant to the icosahedral quasicrystal, J. Phys. Soc. Jpn. {\bf 95}, 044705 (2026). 
\bibitem{WSR} S. Watanabe, Magnetism and topology in Tb-based icosahedral quasicrystal, Sci. Rep. {\bf 11}, 17679 (2021).
\bibitem{Eriksson} F. Eriksson, On the measure of solid angles, Math. Mag. {\bf 63}, 184 (1990). 
\bibitem{Aoyama2021} K. Aoyama and H. Kawamura, Hedgehog-lattice spin texture in classical Heisenberg antiferromagnets on the breathing pyrochlore lattice, Phys. Rev. B {\bf 103}, 014406 (2021). 
\bibitem{Tatara} G. Tatara and H. Kawamura, Chirality-driven anomalous Hall effect in weak coupling regime, J. Phys. Soc. Jpn. {\bf 71}, 2613 (2002).
\bibitem{W2026} S. Watanabe, Ground-state properties of uniaxial magnetic model for antiferromagnetic 1/1 periodic approximants to icosahedral quasicrystals, submitted. 
%
%
\bibitem{WICM2024} S. Watanabe and T. Iwasaki, Effect of magnetic field on whirling-anti-whirling order in icosahedral-quasicrystal approximant, J. Phys. Conf. Ser. {\bf 3161}, 012001 (2026). 
\bibitem{Momma} K. Momma and F. Izumi, Three-dimensional visualization of crystal, volumetric and morphology data, J. Appl. Crystallogr. {\bf 44}, 1272 (2011).
\end{thebibliography}
\end{document}